\documentclass[12pt]{article}           
\usepackage{graphicx, float, amssymb,latexsym}
\def\preprint{}       
\def\finished{} 

\def\title{The Evolution of an Inhomogeneous Universe}

\long\def\abstract{ 
A refined version of a recently introduced method for analysing the 
dynamics of an inhomogeneous irrotational dust universe is presented.
A fully non-perturbative numerical computation of the time dependence of 
volume 
in this framework leads to the following results.
If the initial state of the universe is Einstein-de Sitter with small Gaussian
perturbations, then there is no acceleration even though the  
inhomogeneities strongly affect the evolution. 
A universe with a positive background curvature can exhibit acceleration,
but not in conjunction with reasonable values for the Hubble rate.
Thus the correct values for both quantities can be achieved only
by introducing a positive cosmological constant.
Possible loopholes to this conclusion are discussed;
in particular, acceleration as an illusion created by peculiarities of
light propagation in an inhomogeneous universe is still possible.
Independently of the cosmological constant question, the present formalism
should provide an important tool for precision cosmology.
}

\def\sk{_\mathbf{k}}
\def\Cc#1{S_{#1}}
\def\myexp#1{e^{#1}}

\catcode`\"=\active \let"=\"

\newcommand{\capt}[3]{{\caption{\label{#2} #3}}}
\newcommand{\fref}[1]{Fig.~\ref{#1}}

\def\ifundefined#1{\expandafter\ifx\csname#1\endcsname\relax}
\def\bye{\end{document}}   
\long\def\new#1\endnew{{\bf #1}}
\long\def\del#1\enddel{} 

\def\HS#1 {\hspace*{#1pt}} \def\VS#1 {\vspace*{#1pt}}
\def\BC{\begin{center}}    
\def\EC{\end{center}}

\let\0=\over      \let\1=\vec      \def\2{{1\over2}}    \let\3=\ss
\let\4=\underline \let\5=\overline \let\6=\partial      \def\7#1{{#1}\llap{/}}
\def\8#1{{\textstyle{#1}}}         \def\9#1{{\ifmmode{\pmb{#1}}\else\bf#1\fi}}

          \def\({\left(}       \def\){\right)}

\def\eeql#1 {\label{#1}\eeq}        
\def\beq{\begin{equation}}      \def\eeq{\end{equation}}        
\def\bea{\begin{eqnarray}}      \def\eea{\end{eqnarray}} 

\def\mao#1{\mathop{\rm #1}\nolimits}  
\def\tr{\mao{tr}}

\let\and=\wedge

\let\bra=\langle        \let\ket=\rangle        \def\<#1\>{\bra #1 \ket}

\def\rel#1 #2{\buildrel #1 \over {#2}}  
\def\fnote#1#2{\begingroup\def\thefootnote{#1}\footnote{#2}
                \addtocounter{footnote}{-1}\endgroup}   

          \let\d=\delta   
         \let\th=\theta  
     \let\m=\mu      \let\dh=\vartheta
\let\n=\nu                  \let\r=\rho     \let\s=\sigma 
                 \let\ph=\varphi
               \let\S=\Sigma 
        \let\L=\Lambda     \let\D=\Delta

   \def\cd{{\cal D}}

\def\cm{{\cal M}}    
 \def\cR{{\cal R}}

\def\IR{{\mathbb R}}

\def\plb#1 #2 {Phys. Lett. {\bf B#1} #2 }
\def\phr#1 #2 {Phys. Rep. {\bf  #1} #2 }        
\def\npb#1 #2 {Nucl. Phys. {\bf B#1} #2 }
\def\aph#1 #2 {Ann. Phys. {\bf #1} #2 }         
\def\jmp#1 #2 {J. Math. Phys. {\bf #1} #2 }
\def\jgp#1 #2 {J. Geom. Phys. {\bf #1} #2 }
\def\prd#1 #2 {Phys. Rev. {\bf D#1} #2 }
\def\prl#1 #2 {Phys. Rev. Lett. {\bf #1} #2 }
\def\rmp#1 #2 {Rev. Mod. Phys.  {\bf #1} #2 }
\def\zpc#1 {Z. Phys. {\bf #1C} }
\def\cmp#1 #2 {Commun. Math. Phys. {\bf #1} #2 }
\def\cqg#1 #2 {Class.Quant.Grav. {\bf #1} #2 }
\def\mpl#1 {Mod. Phys. Lett. {\bf A#1} }
\def\cpc#1 {Computer Phys. Commun. {\bf #1} }   
\def\ijmp#1 {Int. J. Mod. Phys. {\bf A#1} }
\def\ijmpC#1 {Int. J. Mod. Phys. {\bf C#1} }
\def\atmp#1 {Adv. Theor. Math. Phys. {\bf #1} }

\def\BP{\begin{picture}} \def\EP{\end{picture}}         
\newcounter{TRefNX} \let\OLDcite=\cite  \makeatletter
\def\makeTRefs#1{\@for  \NewTRef:=#1\do{\global\makeTRef{\NewTRef}}}
\def\makeTRef#1{\ifundefined{TRef#1}\stepcounter{TRefNX}%
\expandafter\xdef\csname TRef#1\endcsname{\theTRefNX}\fi}\makeatother
\def\NEWcite#1{\makeTRefs{#1}\OLDcite{#1}}  

\ifundefined{draftmode} {} 
\else        
   \let\cite=\NEWcite
   \newcount\HOUR  \HOUR=\the\time \divide\HOUR by 60   \multiply \HOUR by 60 
   \newcount\MIN   \MIN=\the\time  \advance\MIN by -\HOUR  \divide\HOUR by 60
   \ifnum  \MIN>9  \def\printTIME{{\it\the\HOUR\,:\,\the\MIN}}
   \else   \def\printTIME{{\it\the\HOUR\,:\,0\the\MIN}} 
   \fi 
   
   \def\LLab#1{\BP(0,0)\unitlength=1mm\put(-12,.5){\makebox(0,0)[cr]{\small #1
        \rlap{$_{_{\makeatletter\csname TRef#1\endcsname\makeatother}}$}}}\EP}
\fi

\begin{document}


{\hfill\preprint }
\vskip 15mm
\begin{center} 
{\LARGE\bf   \title }\vskip 10mm
Harald Skarke\fnote{*}{skarke@hep.itp.tuwien.ac.at}\\[3mm]
Institut f\"ur Theoretische Physik, Technische Universit\"at Wien\\
Wiedner Hauptstra\ss e 8--10, 1040 Wien, Austria
        
\vfill                  
{\bf ABSTRACT } 
\end{center}    
\abstract

\vfill \noindent \preprint\\[5pt] \finished \vspace*{9mm}
\thispagestyle{empty} \newpage
\pagestyle{plain}

\newpage
\setcounter{page}{1}


\section{Introduction}
While data from cosmological observations are reaching unprecedented
levels of precision, their interpretation still rests on the pioneering 
works of Friedmann, Lemaitre, Robertson and Walker (FLRW), who assumed perfect
spatial homogeneity.
This basic framework is refined by linear perturbation theory which 
describes \emph{small} deviations from uniformity very well,
thereby providing an excellent description of the physics of the
early universe.
In the present cosmological era the deviations from homogeneity
are definitely not small.
Therefore a clear procedure for relating measurements from an inhomogeneous 
universe to some FLRW model is needed. 
The fact that this is an open question that needs to be addressed properly 
was recognized already
in Ref.~\cite{EllisS}, where it was called the ``fitting problem''.

This issue became particularly important with the advent of data 
\cite{Riess,Perlm} that,
if interpreted in terms of a naive application of the FLRW models,
indicated the presence of a positive cosmological constant $\L$ or a dark 
energy with similar characteristics.
While the majority of cosmologists seem to agree that this is indeed the
correct interpretation, there have also been many alternative proposals
(see e.g.~\cite{CELU} for a review).
This question was also the motivation for the present work 
(as for its predecessor \cite{iiae}), 
which introduces a formalism that transcends perturbation theory in
its treatment of inhomogeneities. 
The methods presented here should have applications
in many areas of precision cosmology independently of the question of
the cosmological constant.

The basic idea is quite simple:
Consider a large domain $\cd$ in an irrotational dust universe, described
in the synchronous gauge.
Divide $\cd$ into a number of regions that are treated as infinitesimal
in the mathematical framework; actually one should think of these 
regions as small in cosmic terms, but large enough to justify the use of 
the irrotational dust approximation.
Then all one has to do is to follow the evolution of the volume of each 
such region, and to add the contributions to get the volume $V_\cd$ of $\cd$.

The present work is most closely related to an approach that was advanced
by authors such as Kolb et al.~\cite{KMR} and R\"as\"anen  \cite{Ras}
(see Ref.~\cite{Bu07} for a review).
In this approach it is argued that accelerated expansion is a real effect 
in the sense that $\ddot a_\cd > 0$ for $a_\cd = V_\cd^{1/3}$;
this is supposed to take place in a universe containing both collapsing 
and expanding regions, when the latter start to dominate the overall behaviour.
Here we follow these authors by also analysing the evolution of $V_\cd$ in an 
irrotational dust universe.
We differ, however, in terms of the methods that we use.
While Refs.~\cite{KMR,Ras,Bu07} and many others use the ordinary
volume average for obtaining expectation values of scalar quantities,
we use the mass-weighted average of Ref.~\cite{iiae} instead.
This has the advantage that averaging
commutes with taking time derivatives. 
Thereby we can circumvent the use of Buchert's equations \cite{Bu99}
which provide a formalism for treating the corresponding non-commutativity 
in the case of the volume average, but at the expense of technical 
complications that have impeded progress beyond perturbation theory 
up to now.

By refining the proposal of Ref.~\cite{iiae} and using standard 
linear perturbation theory to find the probability distribution for the 
initial values of a basic set of geometric quantities, we will arrive at 
a model with very high predictive power.
A numerical computation then gives the following results.
Inhomogeneities always lead to a strong modification of the volume
evolution.
For the case of a flat background and $\L = 0$ the deceleration parameter is
reduced but still remains positive.
By introducing a positive background curvature it is possible to get 
acceleration, which shows that the mechanism advocated in papers such as
Refs.~\cite{KMR,Ras,iiae} works in principle;
however, it is not possible to get correct values for both the deceleration 
parameter and the Hubble rate at the same time.
As one would expect, the introduction of a positive cosmological constant
can account for these parameters correctly.
Unless one of a small number of rather implausible loopholes is realized,
explaining acceleration as a real effect from inhomogeneity is thus ruled out;
nevertheless it is still possible that light propagation is affected by
inhomogeneity in such a way that acceleration is mimicked without actually
taking place.

The outline of this paper is as follows.
In the next section we present an analysis of the evolution of a local
patch in the universe. We start with the basic setup for an irrotational 
dust universe, proceed with the definition of the local scale factor 
and the mass-weighted average, which are the central concepts of the present
approach, and define rescalings of the basic geometric quantities in such a way 
that their evolution equations become as simple as possible.
In Sec.~\ref{secinval} we make the connection with linear perturbation theory.
By comparing our approach with known results we identify the correct set of 
initial conditions, and by using random matrix theory we arrive at the 
probability distribution for the initial values.
In Sec.~\ref{secresults} we present the results of the computations based
on this model, 
mainly in the form of plots of quantities such as the scale factor
$a_\cd$, the deceleration parameter $q$ or $Ht$ over $t$.
Sec.~\ref{secdisc} contains a discussion 
which includes an analysis of possible loopholes to our conclusions.
An appendix gives some details on numerical aspects of our computations. 

\section{Analysis of the evolution}
\subsection{The Irrotational Dust Universe}
Throughout this paper we model our universe as if it consisted of
friction- and pressureless non-relativistic matter (``dust'') without vorticity.
In this case the most natural choice of coordinate system is 
provided by the synchronous gauge:
every dust particle has constant space coordinates, and
the time coordinate just indicates what a clock comoving with the matter
would show, with the temporal origin set to the time of the big bang.
The spacetime manifold is a cartesian product of the time axis $\IR_+$
and a spacelike manifold $\cm$,
\beq {}^{(4)}\!\cm = \IR_+ \times \cm,  \eeql{spacetime}
and the spacetime metric
\beq {}^{(4)}\! ds^2 = -dt^2 + g_{ij}(t,x) dx^i dx^j \eeql{dustmetric}
is expressed in terms of the time dependent metric $g$ on $\cm$.
It is customary to define the expansion tensor $\th^i_j$ and its trace, 
the scalar expansion rate $\th$, by 
\beq
\th^i_j=\2 g^{ik}\dot g_{kj}, \qquad \th = \th^i_i = {\dot{\sqrt{g}}\0\sqrt{g}},
\eeql{exptens}
where dots denote time derivatives.
The shear is defined as the traceless part of the expansion tensor,
\beq \s^i_j = \th^i_j - {\th\0 3}\d^i_j , \eeql{shear}
hence $\th^i_j\th^j_i = {1\03}\th^2 + 2 \s^2$ with $\s^2 = \2 \s^i_j\s^j_i$.
Similarly we decompose the Ricci tensor that corresponds
to the metric $g$ into its trace (the Ricci scalar $R$) 
and its traceless part $r^i_j$,
\beq R^i_j = {R\0 3}\d^i_j +r^i_j.\eeq
Note that here and elsewhere in this paper geometric quantities such as
$R^i_j$ refer to the spatial 3-geometry unless explicitly indicated 
otherwise.
By using standard formulas of Riemannian geometry, one finds that the time
evolution of the Ricci tensor can be written as 
\beq 
\dot R_{ij} = \th^k_{i|jk} + \th^k_{j|ik} -\th_{ij|kl}g^{kl}-\th_{|ij},
\eeql{Rijev}
where the vertical strokes denote covariant spatial derivatives.

As a consequence of our assumption that the matter consists of irrotational
dust moving along the $(1,0,0,0)$--direction, the energy-momentum tensor 
${}^{(4)}\!T_{\m\n}$ has only one non-vanishing
component, namely the energy density $\rho = {}^{(4)}\!T_{00}$, and the 
covariant conservation of ${}^{(4)}\!T_{\m\n}$ becomes
\beq \dot\rho + \th\rho =0.  \eeql{rhoev}
The following equations represent the $00$-, $0i$- and traceless $ij$-parts
of the Einstein equations:
\bea 
{1\0 3}\th^2 - \s^2 + \2 R - \L &=& 8 \pi G_N \rho, \label{einst00}\\
-2\th_{|i} + 3 \s^j_{i|j} &=& 0, \label{einst0i}\\
\dot\s^i_j + \th \s^i_j + r^i_j &=& 0; \label{sigoev}
\eea
the trace part is obeyed automatically if Eqs.~(\ref{rhoev}) -- (\ref{sigoev})
hold.
Upon splitting Eq.~(\ref{Rijev}) into its trace and traceless part and using
Eqs.~(\ref{shear}) and (\ref{einst0i}) we arrive at the following 
evolution equations for the Ricci scalar and the traceless part of the Ricci
tensor:
\bea \dot R + {2\0 3} \th R  &=& -2\s^i_jr^j_i ,\label{Roev}\\
\dot r^i_j +{2\0 3} \th r^i_j &=& -{5\0 4} \s^i_kr^k_j + {3\0 4} \s^k_jr^i_k 
+
   {1\0 6} \d^i_j \s^k_lr^l_k + {Y^{ki}}_{j|k},\label{rijev} 
\eea
with the last term given by
\beq 
{Y^k}_{ij} = {3\0 4} (\s^k_{i|j}+\s^k_{j|i})-\2g_{ij}{\s^k_{m|}}^m-{\s_{ij|}}^k.
\eeq

\subsection{Different scale factors}
In our analysis of the evolution of an inhomogeneous universe a central
role will be played by a local scale factor that differs both from the
global scale factors that are used for homogeneous universes 
and from the averaged scale factors that are often introduced in the context of
averaging prescriptions. 
Since we need all three types of scale factors and it is important not to
confuse them, we now present each of them.
\begin{itemize}
\item $a_\mathrm{FLRW}(t) = a_\mathrm{LPT}(t)$ is the scale factor associated 
with the FLRW metric 
$g_{ij}^\mathrm{(FLRW)}(t,x) = a_\mathrm{FLRW}^2(t) g_{ij}^\mathrm{(h)}(x)$,
where $g_{ij}^\mathrm{(h)}(x)$ is a homogeneous time-independent metric.
The same scale factor is used to treat perturbations within linear perturbation 
theory (LPT) where $g_{ij}^\mathrm{(h)}(x)$ is modified by some small 
perturbation.
In the case of a flat Einstein-de Sitter universe, which seems to provide a 
very good description of the early universe, $g_{ij}^\mathrm{(h)}(x)=\d_{ij}$
and $a_\mathrm{FLRW} = a_\mathrm{EdS} = \hbox{const} \times t^{2/3}$.
\item $a_\cd(t)=V_\cd^{1/3}(t)$ is the scale factor that characterizes the 
evolution of the volume 
\beq V_\cd=\int_\cd \sqrt{g(x,t)} ~d^3 x \eeq
of a given domain $\cd$. 
It is used to compute the Hubble rate $H_\cd(t) = \dot a_\cd(t) / a_\cd(t)$
and the deceleration parameter $q_\cd(t) = - \ddot a_\cd a_\cd / \dot a_\cd^2$.
\item $a_\mathrm{local}(t,x)$ is the
\emph{local} scale factor that we define as
\beq a_\mathrm{local}(t,x) = \({\hat\rho \0 \rho(t,x)}\)^{1\0 3}
\eeql{lsf}
where $\hat \rho$ is a fixed quantity of dimension mass (e.g.~one solar mass).
This means that our local scale factor is just the side length of a cube of 
mass $\hat \rho$ consisting of material of density $\rho$.
As we will show below, this is equivalent to a different definition given
in Ref.~\cite{iiae}.
Whenever we just write $a(t,x)$ we refer to $a_\mathrm{local}(t,x)$.
\end{itemize}
The connection between $a_\mathrm{local}$ and $a_\cd$ is as follows.
By virtue of Eqs.~(\ref{exptens}) and (\ref{rhoev}), 
\beq {d\0 dt} \(\rho(x,t)\sqrt{g(x,t)}\) = 0 \eeq
and therefore the mass content
\beq m_\cd = \int_\cd \rho(x,t)\sqrt{g(x,t)} ~d^3 x  \eeql{mc}
of any domain $\cd\subset\cm$ is time independent, $\dot m_\cd = 0$.
Hence the mass--weighted $\cd$--average \cite{iiae}
\beq 
\< X\>_\cd (t) = {1\0 m_\cd} \int_\cd X(x,t)\rho(x,t)\sqrt{g(x,t)} ~d^3 x
\eeql{xavg}
of any scalar quantity $X(x,t)$ has the property that averaging 
commutes with taking time derivatives, $\< X\>\dot{}_\cd = \<\dot X\>_\cd$.
Note that this would not hold for a pure volume average which would
therefore require the use of Buchert's formalism \cite{Bu99} for treating time
dependencies.
We can now compute the volume of $\cd$ as
\beq V_\cd = \int_\cd \sqrt{g(x,t)} ~d^3 x = m_\cd \< \rho^{-1}\>_\cd 
= {m_\cd\0 \hat \rho}\< a_\mathrm{local}^3\>_\cd ,\eeq
and the scale factor corresponding to the domain $\cd$ as $a_\cd = V_\cd^{1/3}$.

\subsection{Evolution of rescaled quantities}
The evolution equation (\ref{rhoev}) for the density $\rho$ and 
the definition (\ref{lsf}) of the local scale factor imply that the scalar 
expansion rate $\th$ can be expressed as
\beq  \th(t,x) = - {\dot \r(t,x) \0  \r(t,x)} = 
3 {\dot a(t,x) \0  a(t,x)} \eeq
(in Ref.~\cite{iiae} $a$ was defined as the solution of this equation; this
is equivalent to the present definition as given in Eq.~(\ref{lsf})).
Therefore the rescaled quantities
\beq \hat\rho = a^3 \rho,  ~~~ \hat\s^i_j = a^3 \s^i_j,  ~~~ \hat R = a^2 R,
   ~~~ \hat r^i_j = a^2 r^i_j \eeq
obey the simpler evolution equations
\beq
\dot{\hat\rho} = 0,\quad
{\dot{\hat\s}^i_j} = - a \hat r^i_j,\quad
\dot{\hat R} = -2 a^{-3} \hat\s^i_j \hat r^j_i,
\eeql{eveq}
\beq \dot{\hat r^i_j} =   
   a^{-3}\(-{5\0 4} \hat\s^i_k\hat r^k_j +  {3\0 4} \hat\s^k_j\hat r^i_k 
   + {1\0 6} \d^i_j \hat\s^k_l\hat r^l_k\) + a^2{Y^{ki}}_{j|k}.\label{rhatev}
\eeq
In terms of the new rescaled quantities, Eq.~(\ref{einst00}) becomes
an evolution equation for the local scale factor $a$,
\beq 3{\dot a^2 \0a^2} = \hat\s^2 a^{-6} + 8\pi G_N\hat\rho \,a^{-3}
- {1\0 2}\hat R\,a^{-2} +\L.   \label{expev}\eeq
As a consequence of Eqs.~(\ref{eveq}) we can compute the evolution
of $\hat R$ from some initial time $t_{\mathrm{in}}$ onwards as
\del
\beq \hat\s^2(t) = 
\hat\s^2(t_{\mathrm{in}})-\int_{t_{\mathrm{in}}}^ta(\tilde t)\,\hat\s^i_j(\tilde t)
\hat r^j_i(\tilde t)\,d\tilde t, \quad
\eeq
\beq {a^{-2}\0 3}\int_{t_{\mathrm{in}}}^t\({1\0 a^3(\tilde t)}
-{a(\tilde t)\0 a^4(t)}\)
\hat\s^i_j(\tilde t) \hat r^j_i(\tilde t)\,d\tilde t   \eeq
\enddel
\bea
\hat R(t) &=& 
\hat R(t_{\mathrm{in}})-2\int_{t_{\mathrm{in}}}^ta^{-3}(\tilde t)\,\hat\s^i_j(\tilde t)
\hat r^j_i(\tilde t)\,d\tilde t \\
&=& 
\hat R(t_{\mathrm{in}})+2\int_{t_{\mathrm{in}}}^ta^{-4}(\tilde t)\,\hat\s^i_j(\tilde t)
\dot{\hat \s^j_i}(\tilde t)\,d\tilde t\\
&=& \hat R(t_{\mathrm{in}}) 
+ 2 a^{-4}(t)\,\hat\s^2(t) - 2 a^{-4}(t_{\mathrm{in}})\,\hat\s^2(t_{\mathrm{in}})
+{8\0 3}\int_{t_{\mathrm{in}}}^t
\th(\tilde t)a^{-4}(\tilde t)\,\hat\s^2(\tilde t)\,d\tilde t,
\eea
whereby the evolution equation for the local scale factor becomes
\beq 
\dot a^2  = {1\0 3}\hat\s^2_\mathrm{in} a^{-4} + {8\0 3}\pi G_N\hat\rho \,a^{-1}
- {1\0 6}\hat R_\mathrm{in} + {1\0 3}\L\,a^2 - {4\0 9}
\int_{t_{\mathrm{in}}}^t\th(\tilde t)a^{-4}(\tilde t)\,\hat\s^2(\tilde t)\,d\tilde t.
\label{lsfev}\eeq
Note that each term on the right hand side, except for the last, is a Laurent 
monomial in $a$ with a time independent coefficient.
The last term is negative during expansion ($\th > 0$) and positive
during contraction ($\th < 0$), i.e.~its effect is always like that of an
attractive force.

\section{Initial values from linear perturbation theory}\label{secinval}
The next step is to consider which initial values $a_\mathrm{in}$, 
$\hat\s_\mathrm{in}$ etc.~should be used in Eq.~(\ref{lsfev}).
It is generally accepted that linear perturbation theory provides an 
excellent description of the evolution of the early universe with all of
its inhomogeneities, so this is what we are going to use.

\subsection{Linear perturbation theory}

The application of linear perturbation theory to 
an irrotational dust universe in the synchronous gauge is analysed in 
detail in Ref.~\cite{LS}.
There, the initial scalar perturbations are parametrized in terms of several 
scalar Gaussian random fields.
Upon taking into account relations between these fields and ignoring
decaying modes, the result is that the relevant contributions
all come from a single time-independent function $C(x)$. 
For a flat background, where 
$a_\mathrm{FLRW} = a_\mathrm{EdS} = \hbox{const} \times t^{2/3}$,
the corresponding linearly perturbed metric is
\beq g_{ij}^\mathrm{(LPT)}(t,x) = a_\mathrm{FLRW}^2(t)
  \(\d_{ij} + {10\0 9}{a_\mathrm{FLRW}^2\0 t^{4\0 3}}C(x)\d_{ij} 
   + t^{2\0 3} \partial_i\partial_jC(x)\).   \eeq
A straightforward calculation results in the first order expressions
\bea R^i_j(t,x) &=& -  {5\0 9}t^{-{4\0 3}}( \d^{ik}\partial_k\partial_j + 
   \d^i_j\d^{kl}\partial_k\partial_l)C(x),\\
\th^i_j(t,x) &=& {2\0 3}t^{-1}\d^i_j 
   + {1\0 3}t^{-{1 \0 3}}\d^{ik}\partial_k\partial_jC(x)
\eea
for the Ricci tensor and the expansion tensor, respectively.
These quantities depend on $C(x)$ only via the symmetric matrix 
\beq \partial_i\partial_jC(x)=\Cc{ij}(x)=s_{ij}(x)+{1\0 3}\d_{ij}S(x)\eeq
of second spatial derivatives; $S$ and $s_{ij}$ are the trace and
traceless part of $\Cc{ij}$.
Upon decomposing $R^i_j$ and $\th^i_j$ we get
\bea
R &=& -  {20\0 9}t^{-{4\0 3}}S,\\
\th &=& 2t^{-1} + {1\0 3}t^{-{1 \0 3}}S, \\
r^i_j &=& -  {5\0 9}t^{-{4\0 3}}\d^{ik}s_{kj},\\
\s^i_j &=& {1\0 3}t^{-{1 \0 3}}\d^{ik}s_{kj}.\label{siglpt}
\eea
It is easily checked that these quantities satisfy Eqs.~(\ref{einst0i}), 
(\ref{sigoev}) and (\ref{Roev}) at first order.
Furthermore, with Eq.~(\ref{einst00}) we get
\beq 6\pi G_N \rho = t^{-2} - \2 t^{-{4\0 3}}S,\eeq
which corresponds to density perturbations of
\beq {\D\r \0 \r} = - \2 t^{2\0 3} S \eeql{Deltarho}
in the early universe.

\subsection{The initial value problem}\label{ivp}
We now want to use the results from linear perturbation theory to find the 
initial values for the set of differential equations describing the evolution.
We start with considering the  scaling properties of our rescaled (``hatted'') 
quantities $\hat R$ etc.~under $t\to 0$, where $a\sim t^{2/3}$.
Eq.~(\ref{siglpt}) shows that $\hat \s = a^3 \s$ is proportional 
to $t^{5/3}$ near $t=0$.
\del
Hence, upon choosing $t_\mathrm{in}=0$, the term $\hat\s^2_\mathrm{in}a^{-4}/3$ in
Eq.~(\ref{lsfev}) vanishes and the integral 
$\int_{t_{\mathrm{in}}}^t\th(\tilde t)a^{-4}(\tilde t)\,\hat\s^2(\tilde t)\,d\tilde t$
is proportional to $t^{2/3}$ near $t=0$.
So the right hand side of Eq.~(\ref{lsfev}) 
\enddel
Applying this fact to Eq.~(\ref{lsfev}) with the choice of 
$t_\mathrm{in} = 0$, we find that $\hat\s^2_\mathrm{in}a^{-4}/3=0$  
and that the integral 
$\int_0^t\th(\tilde t)a^{-4}(\tilde t)\,\hat\s^2(\tilde t)\,d\tilde t$
is well-defined and proportional to $t^{2/3}$ near $t=0$.
Hence the right-hand side of Eq.~(\ref{lsfev}) is dominated by the term 
${8\0 3}\pi G_N\hat\rho \,a^{-1}$, implying
\beq \lim_{t\to 0}\,{a\0 t^{2\0 3}} = (6\pi G_N\hat \rho)^{1/3}. \eeq
Therefore we have the identifications
\bea
\hat R_{\mathrm{in}}(x) &=& -  
{20\0 9}(6\pi G_N\hat \rho)^{2\0 3}S(x),\\
({\hat r}_{\mathrm{in}})^i_j(x) &=& 
-  {5\0 9}(6\pi G_N\hat \rho)^{2\0 3}\d^{ik}s_{kj}(x).
\eea
Thus our rescaling has led to quantities that are finite
at the origin, providing us a with a well defined initial value problem.
It is useful to perform one more set of redefinitions in order to get
dimensionless variables. 
The entries of the matrix $\Cc{ij}$ must have dimensionality $t^{-2/3}$.
We can use any quantity $U$ of the same dimensionality 
to make a transformation to 
\bea
\bar \Cc{ij} &=& U^{-1}\Cc{ij}, \\
{\bar t} &=& U^{{3\0 2}} t,\label{ttrafo}\\
\bar a &=& (6\pi G_N\hat \rho)^{-{1\0 3}} Ua, \label{atrafo}\\
\bar R &=& (6\pi G_N\hat \rho)^{-{2\0 3}} U^{-1}\hat R,\\
\bar r &=& (6\pi G_N\hat \rho)^{-{2\0 3}} U^{-1}\hat r,\\
\bar \s &=& (6\pi G_N\hat \rho)^{-1}U^{{3\0 2}}\hat \s,\\
\bar\L &=& U^{-3}\L,
\eea
so that
\bea
\bar R_{\mathrm{in}}(x) &=& - {20\0 9}\bar S(x),\label{Rbarin}\\
({\bar r}_{\mathrm{in}})^i_j(x) &=& 
-  {5\0 9}\d^{ik}\bar s_{kj}(x),\\
({\bar \s}_{\mathrm{in}})^i_j(x) &=& 0.\label{sigbarin}
\eea
In terms of these dimensionless quantities Eq.~(\ref{lsfev}) becomes
\beq {d\0 d{\bar t}}({\bar a}^{3\0 2}) = 
  \pm \sqrt{1 - {3\0 8}\bar a  \bar R_{\mathrm{in}} +{3\0 4}\bar a^3\bar\L
 - 3 \bar a \int_0^{\bar t} {d\bar a\0 d{\bar t'}} \,{\bar a}^{-5}\bar\s^2 
  d{\bar t'}}. 
\eeql{a32ev}
Away from small values of $\bar a$
it is useful to take a further derivative to arrive at the simpler equation
\beq 
{9\0 2} {d^2\bar a\0 d{\bar t}^2} = 
  - \bar a^{-2} + {3\0 2} \bar a\bar\L - 3\bar a^{-5} {\bar\s}^2. 
\eeq
Either of these equations for $\bar a$ must be supplemented by the evolution 
equation
\beq {d{\bar\s}^i_j\0 d{\bar t}} = - \bar a \bar r^i_j \eeq
for $\bar \s$.
Finally $\bar r^i_j$ must be modelled. 
Here we have to depart from the exact description provided by 
Eq.~(\ref{rhatev}) because 
of the last term ${Y^{ki}}_{j|k}$.
The following scenarios with rising level of complexity and precision
appear natural.
\begin{enumerate}
\item $\bar r^i_j=0$: Then we also have $\bar \s = 0$.
This is essentially the model proposed in \cite{iiae}.
\item $\bar r^i_j=(\bar r_{\mathrm{in}})^i_j$: Then 
${\bar\s}^i_j({\bar t}) = - (\bar r_{\mathrm{in}})^i_j \bar A({\bar t})$
where $\bar A({\bar t})  
=\int_0^{\bar t}\bar a({\bar t'})d{\bar t'}$, and 
$\bar \s^2 = \bar A^2 \bar r^2_{\mathrm{in}}$.
\del
\beq 
\bar \s^2 = \bar A^2 \bar r^2_{\mathrm{in}}   = {25\0 243}\bar A^2\sin^2\th,
\eeq
and the evolution of $\bar a$ is described by
\beq 
{d\0 d{\bar t}}({\bar a}^{3\0 2}) = \pm \sqrt{1 + {5\0 6} \sqrt{5}\cos\th\, \bar a  
- {25\0 81}\sin^2\th\,\bar a \int_0^{\bar t} {d\bar a\0 d\tilde{\bar t}} 
\,{\bar a}^{-5}\bar A^2 d\tilde{\bar t}}
\eeq
or
\beq 
{9\0 2} {d^2\bar a\0 d{\bar t}^2} = 
- \bar a^{-2} - {25\0 81}\sin^2\th\,\bar a^{-5} {\bar A}^2.
\eeq
\del
where 
\beq c_1 = {5\0 6} \sqrt{5}\cos\th,\qquad 
c_2 = {25\0 81}\sin^2\th.  \eeq
\enddel
\item $\bar r^i_j$ non-constant and modelled by 
\beq {d{\bar r^i_j}\0 d\bar t} =   
   \bar a^{-3}\(-{5\0 4} \bar\s^i_k\bar r^k_j +  {3\0 4} \bar\s^k_j\bar r^i_k 
   + {1\0 6} \d^i_j \bar\s^k_l\bar r^l_k\).
\eeql{rbarijev}
Here the only aberration from an exact description stems from the omission
of the $Y$-term in Eq.~(\ref{rhatev}).
Note that if we start in a coordinate system in which 
$(\bar r_{\mathrm{in}})^i_j$ is diagonal then both $\bar r^i_j$ and $\bar\s^i_j$
remain diagonal in that system.
\item Exact description using the full Eq.~(\ref{rhatev}).
This is beyond the present study  since we have no
handle on the $Y$-term.
\end{enumerate}
None of the equations from (\ref{Rbarin}) onwards contains $U$ 
explicitly.
Let us denote the solutions to these equations with initial values 
$\bar \Cc{ij}$ by $\bar a(\bar t; \bar \Cc{ij}, \bar \L)$.
By comparing the results for different normalization factors $U$ and 
$U' = q U$ we deduce that
\beq 
q \bar a(\bar t; \bar \Cc{ij}, \bar \L) = 
\bar a(q^{3\0 2}\bar t; q^{-1}\bar \Cc{ij}, q^{-3}\bar \L).
\eeql{qU}

\subsection{The probability distribution}\label{subsecpd}
Our next aim is to embed the evolution equations into a statistical model 
for the inhomogeneity of the universe.
As a first step we now 
want to find the distribution of the eigenvalues of the matrix $S_{ij}$.
Using the Fourier decomposition 
\beq 
C(x) = \int(a\sk + i b\sk) e^{i\,\mathbf{k}\cdot \mathbf{x}}{d^3 k\0(2\pi)^{3\0 2}},
\qquad a_{-\mathbf{k}} = a\sk, ~~ b_{-\mathbf{k}} = -b\sk,
\eeq
we obtain
\beq \Cc{ij}(0) = {\partial^2C(x)\0\partial x_i\partial x_j}|_{x=0} 
= -\int a\sk k_i k_j {d^3 k\0(2\pi)^{3\0 2}}.\eeq
Since $C(x)$ is a Gaussian random field, the 
probability distributions for modes $a\sk$ and $a_{\tilde\mathbf{k}}$ are 
independent unless $\mathbf{k}=\tilde\mathbf{k}$.
Because of translational invariance we can compute the variances 
$\<\Cc{ij}^2\>$ and covariances $\<\Cc{ij}\, \Cc{lm}\>$ of the elements of 
the matrix $\Cc{ij}(x)$ 
at $x=0$:
\beq 
\<\Cc{ij}\, \Cc{lm}\> 
= \<\int a\sk a_{\tilde\mathbf{k}}\, k_ik_j  \, \tilde k_l\tilde k_m \,
  {d^3 k \,d^3 \tilde k \0 (2\pi)^3}\>
= I \times \int_{S^2} e_ie_je_le_m dA.
\eeql{Iint}
The second step involved using the independence of different modes to
get $\d(\mathbf{k}-\tilde\mathbf{k})$, writing $k_i = e_i |\mathbf{k}|$ and
splitting off an integral over the 
unit sphere $S^2 = \{\mathbf{e} = \mathbf{k} / | \mathbf{k} | \}$ with area 
element $dA$ (this is allowed by the rotational invariance of the $a\sk$);
the constant $I$ represents the result of integrating over $|\mathbf{k}|$
and taking the expectation value $\<~\ldots~\>$ in the Gaussian distribution.
Then one easily finds (e.g.~by using polar coordinates)
\bea
&&\<\Cc{11}^2\> 
= {4\0 5}I\pi,\label{vars}\\
&&\<\Cc{12}^2\> = \<\Cc{11}\,\Cc{22}\>
= {4\0 15}I\pi,\\
&&\<\Cc{11}\,\Cc{12}\> = \<\Cc{11}\,\Cc{23}\> = \<\Cc{12}\,\Cc{13}\> = 0,
\label{covars}\eea
with the same results for expressions that can be obtained by permutations
of the labels 1, 2, 3.

The matrix elements $\Cc{ij}$, being derivatives of the Gaussian random field
$C$, must themselves obey a Gaussian distribution;
moreover, this distribution must be invariant under orthogonal conjugation.
There exists a well-developed theory of Gaussian random matrices 
(see e.g.~Ref.~\cite{Akemann}; a useful brief summary is provided by Wikipedia
\cite{Wikirm}).
In particular, a symmetric invariant Gaussian random matrix
should be proportional to $M + \nu \mathbf 1$, where $M$ is a 
matrix drawn from a Gaussian orthogonal ensemble, $\nu$ is a Gaussian random 
variable and $\mathbf 1$ represents the unit matrix.
A Gaussian orthogonal ensemble is the set of symmetric 
$n\times n$ matrices $M$ equipped with the probability density 
$\exp(-{n\0 4}\tr M^2)$, i.e.~the matrix elements of $M$ are uncorrelated 
with variances of $1/n$ and $2/n$ for the off-diagonal and diagonal cases, 
respectively.
The eigenvalues $\m_i$ of $M$ are then distributed according to the density
\beq 
p(\m_1,\ldots,\m_n) \sim \myexp{-{n\0 4}\sum_{i=1}^n \m_i^2}
\prod_{1\le i <  j \le n}|\m_i - \m_j|. 
\eeq
In the present case of $n=3$ we can reproduce the results of Eqs.~(\ref{vars})
to (\ref{covars}) 
by choosing $\nu$ to have variance $1/3$, and 
\beq 
\bar \Cc{ij} = U^{-1} \Cc{ij} = M_{ij} + \nu \d_{ij} 
~~\hbox{ with } ~~U=\sqrt{{4 \0 5}I\pi}.
\eeq
The resulting density
\beq
p(\nu, \m_1, \m_2, \m_3) \sim 
\myexp{-{3\0 4}(\m_1^2 + \m_2^2 + \m_3^2 + 2 \nu^2)}
|(\m_1 - \m_2)(\m_1 - \m_3)(\m_2 - \m_3)|
\eeq
can be transformed via 
$\S = \m_1+\m_2+\m_3$, $\bar S = 3 \nu + \S$, $\bar \d_i=\m_i-{\S\0 3}$ to
\beq
p(\bar S, \S, \bar \d_1, \bar \d_2) \sim 
\myexp{-{1\0 10}\bar S^2-{5\0 12}(\S -{2\0 5}\bar S)^2-{3\0 4} 
(\bar \d_1^2 + \bar \d_2^2 + \bar \d_3^2)}
|(\bar \d_1-\bar \d_2)(\bar \d_1-\bar \d_3)(\bar \d_2-\bar \d_3)|,
\eeq
where $\bar \d_1$, $\bar \d_2$ and $\bar \d_3= - \bar \d_1 - \bar \d_2$ 
are the eigenvalues of $\bar s_{ij}$.
Since $\S$ plays no role in our further computations it can be integrated
out. 
With one more change of variables such that
\beq 
\quad \bar \d_1 = {2\0 3} \bar \d\cos\ph,
\quad \bar \d_2 = {2\0 3} \bar \d\cos(\ph + {2\pi\0 3}),
\quad \bar \d_3 = {2\0 3} \bar \d\cos(\ph + {4\pi\0 3}),
\eeql{deltabar}
the probability density can be written as
\beq 
p(\bar S, \bar \d, \ph) \sim 
\myexp{-{1\0 10}(\bar S-\bar S_b)^2-{1\0 2}\bar \d^2}\bar \d^4\, 
|\sin(3 \ph)|,   
\eeql{probmeas}
where we have included the possibility of a nonzero background curvature
by introducing a background value $S_b$ for $S$.

\section{Results}\label{secresults}

Putting the results of the previous sections together we now have all
the ingredients required to compute the volume of the domain $\cd$:
this is achieved by integrating the 
local volumes with the measure (\ref{probmeas}):
\beq
V_\cd(\bar t;\bar S_b,\bar \L)  
\sim \int \bar a^3({\bar t};\bar S,\bar \d,\ph, \bar \L) 
\myexp{-{1\0 10}(\bar S-\bar S_b)^2-{1\0 2}\bar \d^2}\bar \d^4\, 
    \sin(3 \ph) d\bar S\,d\bar \d\, d\ph.
\eeql{result}
Here $\bar a({\bar t};\bar S,\bar \d,\ph, \bar \L)$ is the solution of 
Eqs.~(\ref{a32ev}) to (\ref{rbarijev}) with initial values from 
Eqs.~(\ref{Rbarin}) to (\ref{sigbarin}), where $\bar s$ is taken to be the
diagonal matrix whose eigenvalues are given in Eq.~(\ref{deltabar}).
This computation was performed numerically, with details given in the appendix.

Let us now present the results.
We continue using the dimensionless variables $\bar S$, $\bar t$ and
$\bar \L$ that
were introduced in Sec.~\ref{ivp}, but drop the bars henceforth.

\subsection{Vanishing cosmological constant}

\begin{figure}[H]
\begin{center}
\includegraphics[width=12cm]{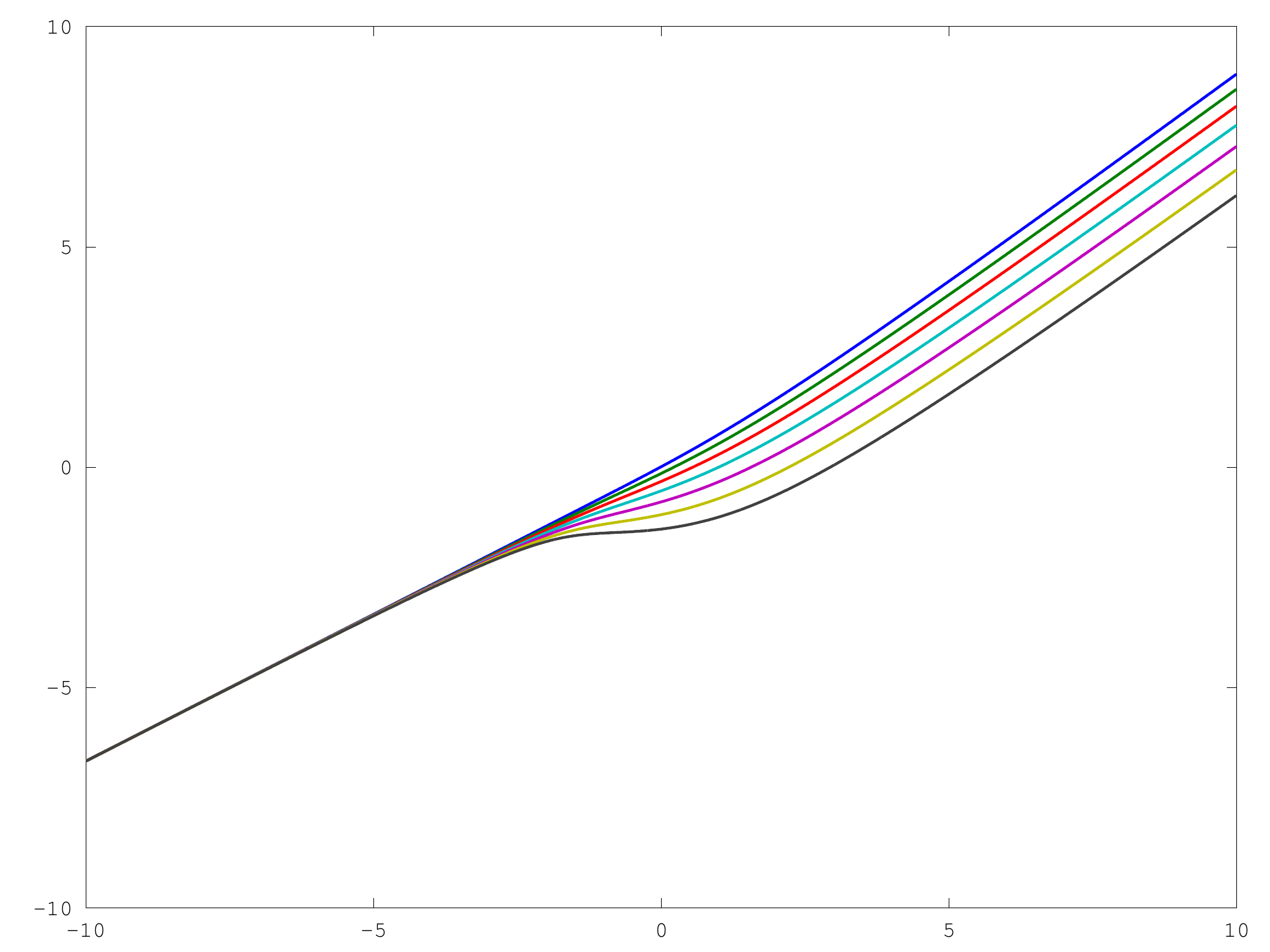}
\capt{12cm}{beu7la}{ln($a_\cd$) over ln($t$) for $S_b = 0/-1/-2/-3/-4/-5/-6$}
\end{center}
\end{figure}

We start with the case of $\L = 0$, where we consider the possibility of
having a non-vanishing background curvature $R_b$ determined via 
Eq.~(\ref{Rbarin}) from $S_b$; note that negative $S_b$ corresponds
to positive curvature and vice versa.

\fref{beu7la} shows a plot of ln($a_\cd$) over ln($t$) for 
$S_b = 0/-1/-2/-3/-4/-5/-6$.
The fact that $V_\cd = a_\cd^3 \sim t^2$ for small $t$ manifests itself by
each of the lines starting with a slope of 2/3;
an EdS universe would correspond to the case where the same linear
relationship between ln($a_\cd$) and ln($t$) would hold everywhere.
Here and elsewhere we have chosen the convention of normalizing $a_\cd$ in
such a way that $a_\cd^3/t^2\to 1$ for $t\to 0$;
hence in the plot the straight line 
corresponding to EdS would pass through the origin of the plot.
The effect of inhomogeneity on the case with $S_b = 0$ (the highest, blue line)
is just to change the slope somewhat.
For the cases of $S_b < 0$ the impact of having inhomogeneities is much more
pronounced: without them, these universes would recollapse, but in their 
presence the overall expansion is just reduced while the regions that start 
with positive curvature collapse, and picks up speed again once the 
collapse is over.
Of course our methods also work for $S_b>0$ ($R_b < 0$) but the resulting 
universes are not very interesting since they look more or less like
open FLRW universes.

\begin{figure}[H]
\begin{center}
\includegraphics[width=12cm]{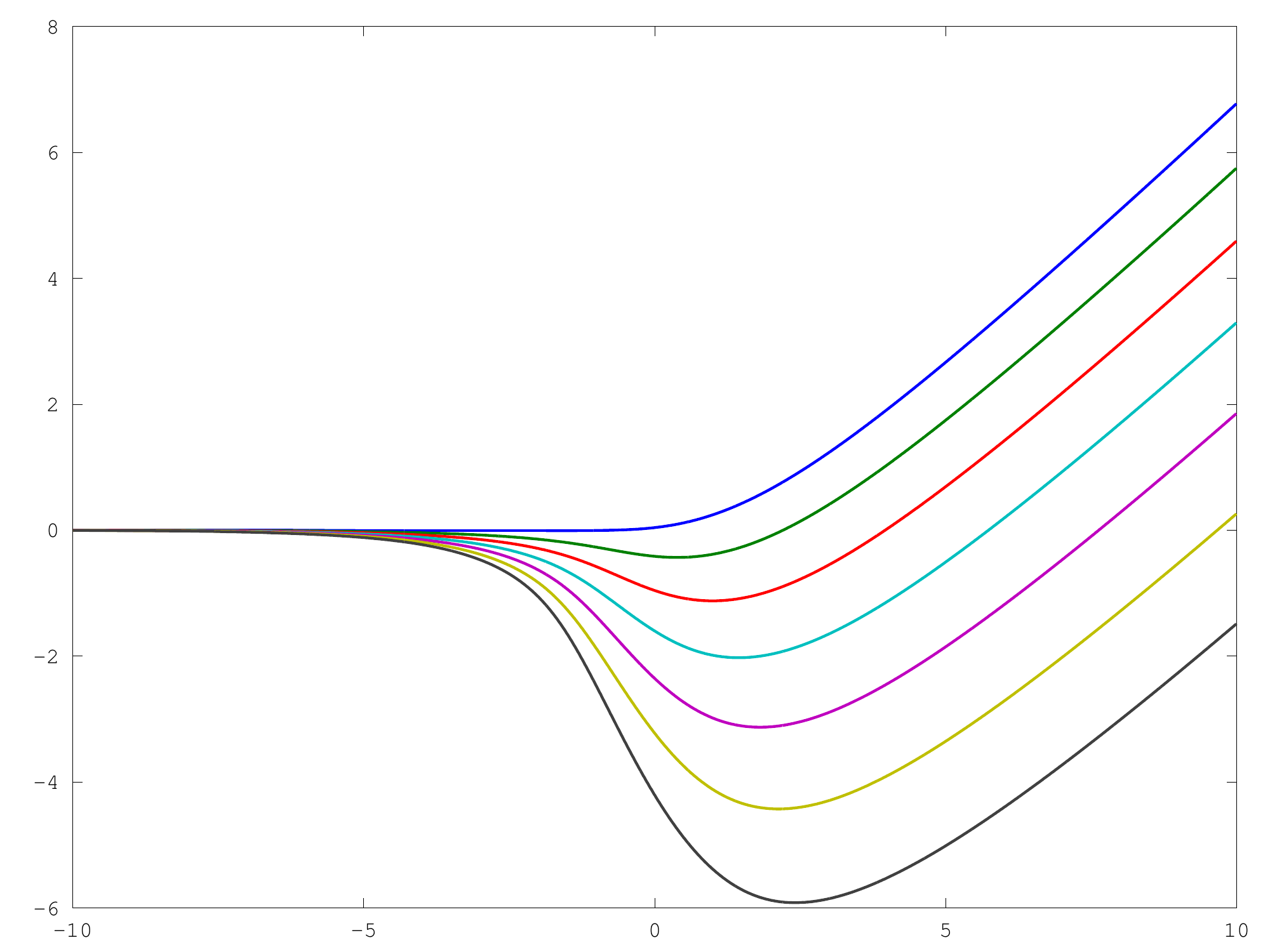}
\capt{12cm}{beu7lvot2}{ln($V_\cd/t^2$) over ln($t$) for $S_b = 0/-1/-2/-3/-4/-5/-6$}
\end{center}
\end{figure}

The next plot (\fref{beu7lvot2}), which diplays ln($a_\cd^3/t^2$) over 
ln($t$), is just a linearly transformed version of the first one, which 
perhaps gives a better idea of how the inhomogeneities modify the 
evolution. 
Here the EdS case would correspond to a horizontal line. 
For the cases with nonzero background, one can clearly see the onset of 
collapse before the minority of expanding regions starts to control the
overall behaviour.

\begin{figure}[H]
\begin{center}
\includegraphics[width=12cm]{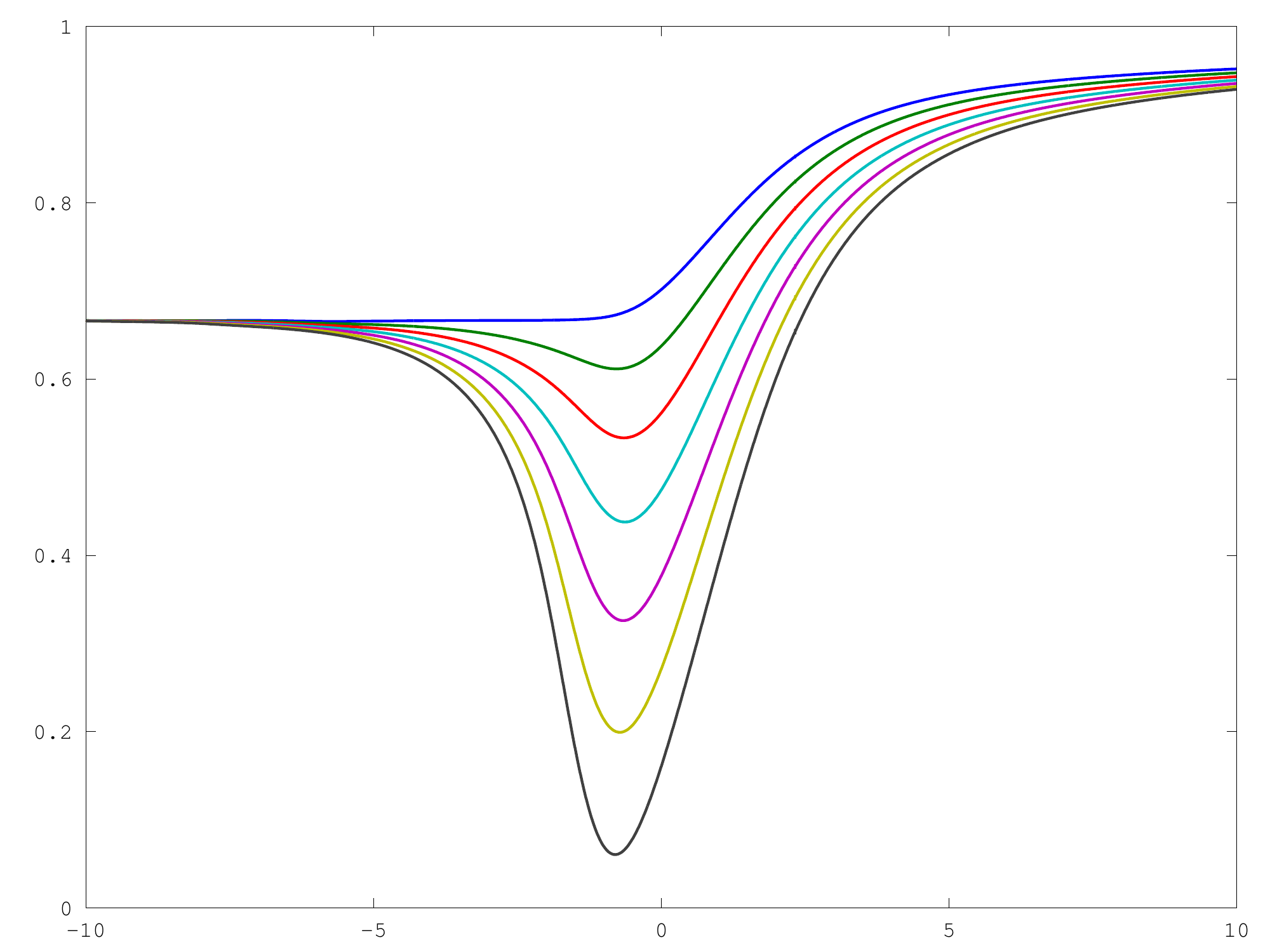}
\capt{12cm}{beu7Ht}{$Ht$ over ln($t$) for $S_b = 0/-1/-2/-3/-4/-5/-6$}
\end{center}
\end{figure}

In \fref{beu7Ht} we see the graph of $Ht$ (with $H$ being, of course,
$H_\cd = \dot a_\cd /a_\cd$).
Here we start again with the EdS value, which is $Ht = 2/3$, and seem to 
converge 
very slowly towards the open FLRW case of $Ht=1$. 
The behaviour around $t\approx 1$ depends strongly upon the value of the 
background curvature: the higher $|S_b|$, the lower $Ht$ can go, and 
indeed for the case of $S_b = -7$ (not shown in the plot) we would get 
a phase of negative $\dot a_\cd$, hence negative $Ht$.

\begin{figure}[H]
\begin{center}
\includegraphics[width=12cm]{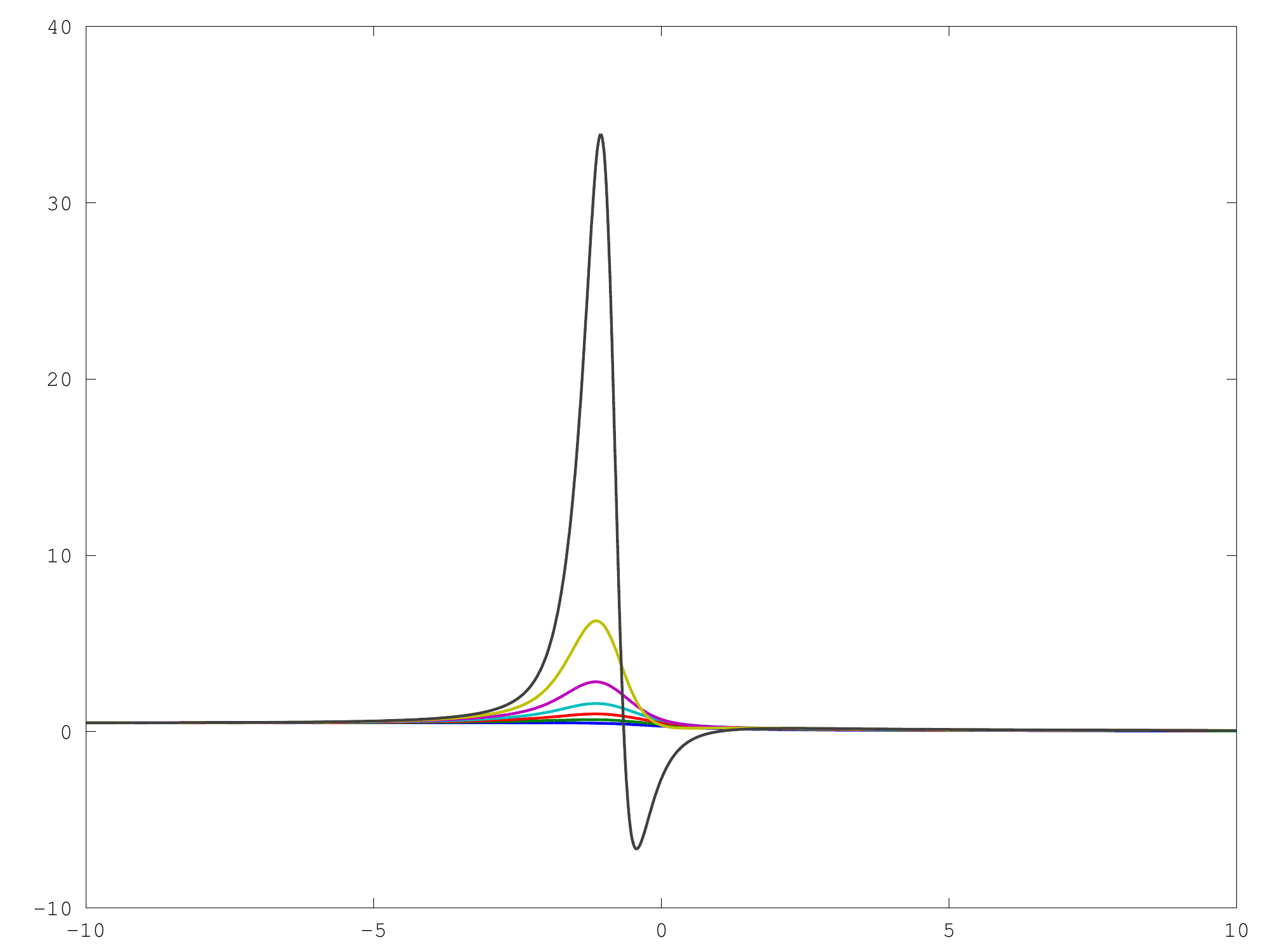}
\capt{12cm}{beu7decel}{Deceleration parameter over ln($t$) for $S_b = 0/-1/-2/-3/-4/-5/-6$}
\end{center}
\end{figure}

\begin{figure}[H]
\begin{center}
\includegraphics[width=12cm]{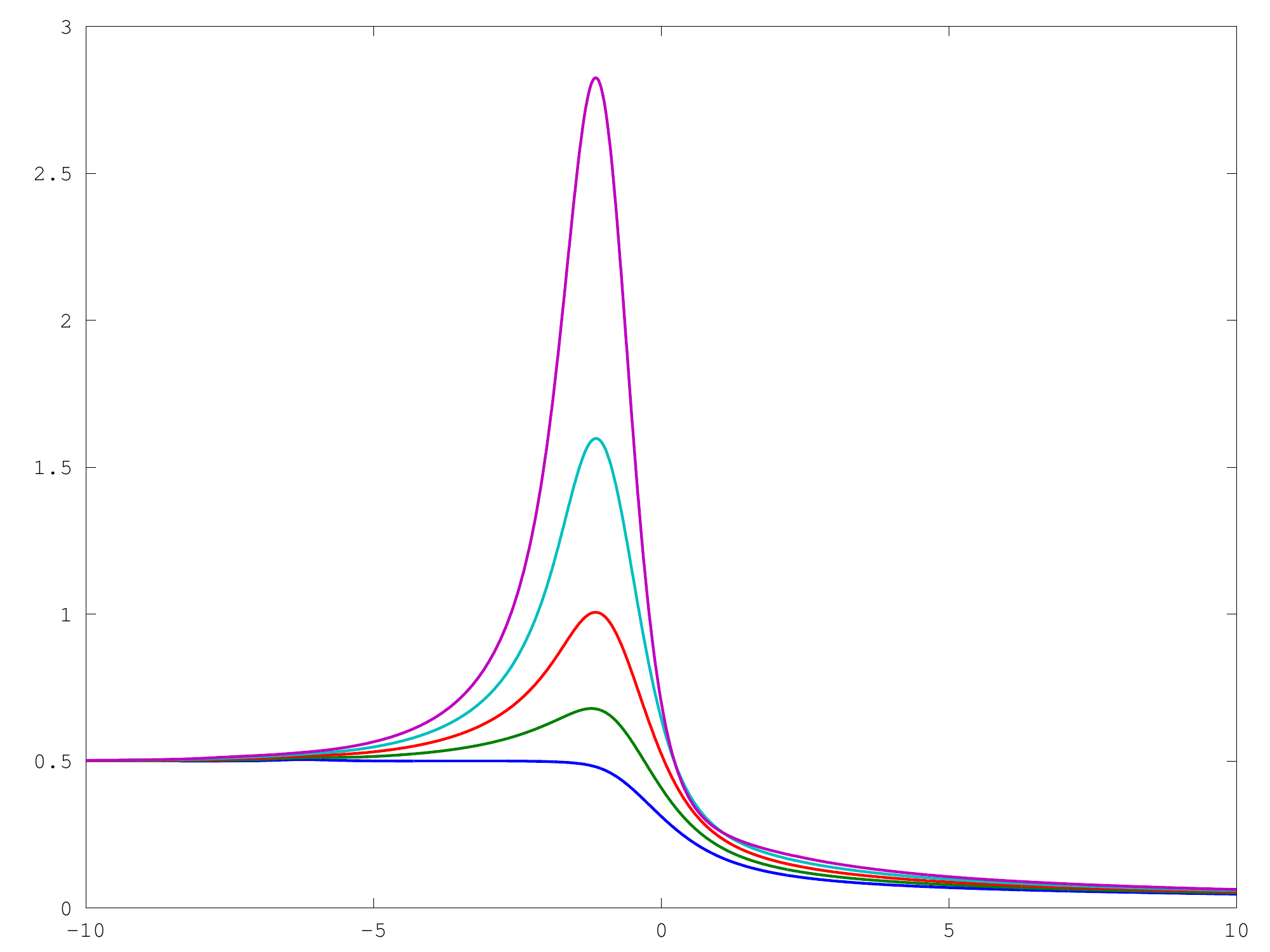}
\capt{12cm}{beu7deceltom4}{Deceleration parameter over ln($t$) for 
$S_b = 0/-1/-2/-3/-4$}
\end{center}
\end{figure}

\fref{beu7decel} and \fref{beu7deceltom4} display the deceleration parameter 
$q(t) = - \ddot a_\cd a_\cd / \dot a_\cd^2$ over ln($t$).
As \fref{beu7decel} shows, the deceleration parameter can become significantly
negative if the background curvature is large enough.
In fact, $q$ diverges for higher values of $|S_b|$ since $\dot a_\cd$ will
then pass through zero.
The asymptotic behaviour for $\ln(t)\to\pm\infty$ is easier to see in 
\fref{beu7deceltom4}: again, near $t=0$ the EdS behaviour (with $q=1/2$)
is approached, whereas for $t\to\infty$ a value of $q=0$ is approached, 
just like for an open universe.

\begin{figure}[H]
\begin{center}
\includegraphics[width=12cm]{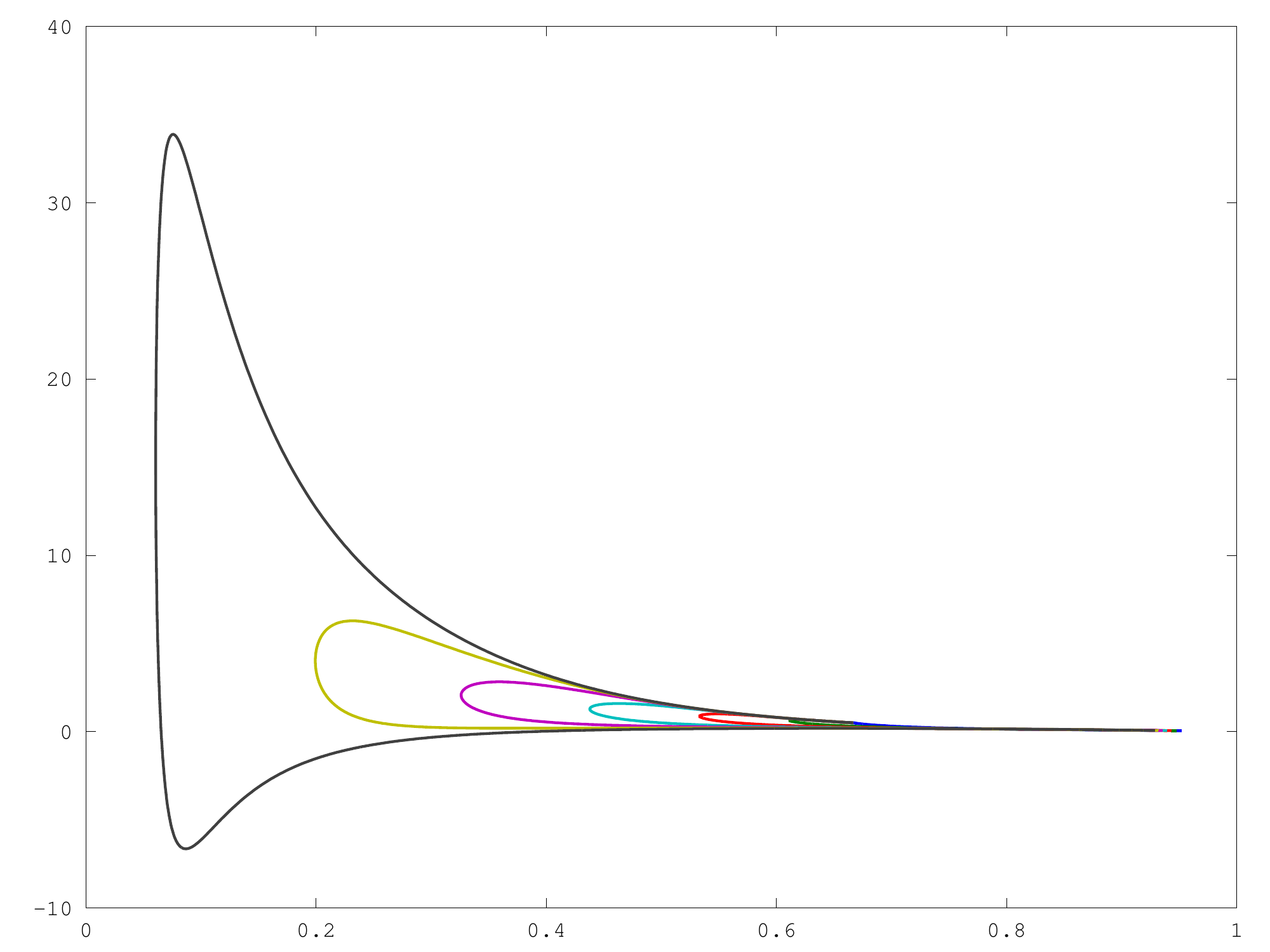}
\capt{12cm}{beu7qHt}{Deceleration parameter over $Ht$ for 
$S_b = 0/-1/-2/-3/-4/-5/-6$}
\end{center}
\end{figure}

\fref{beu7qHt} combines data from the previous plots in a different way. 
Now the deceleration parameter $q$ is displayed over $Ht$.

\begin{figure}[H]
\begin{center}
\includegraphics[width=12cm]{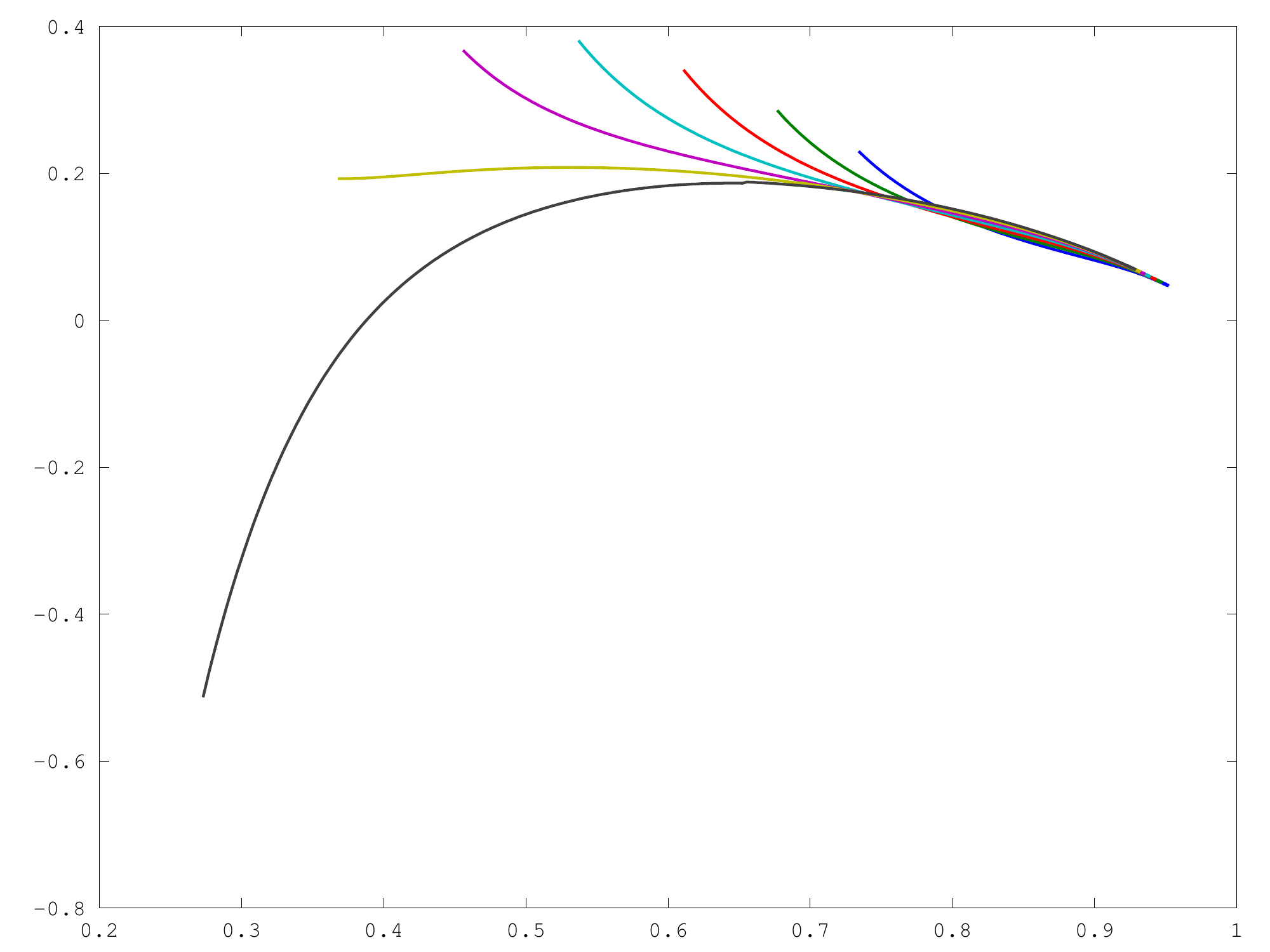}
\capt{12cm}{beu7qHtexc}{Deceleration parameter over $Ht$ for 
$S_b = 0/-1/-2/-3/-4/-5/-6$ (only $\ln(t)\geq 1/2$)}
\end{center}
\end{figure}

\fref{beu7qHtexc} shows the part of \fref{beu7qHt} that corresponds only
to the values $\ln(t)\geq 1/2$, with a strongly expanded scale in the
vertical direction corresponding to $q$.
It is fairly clear from these plots that $Ht\approx 1$ and $q\approx -1/2$ 
cannot be achieved simultaneously by the present model with $\L = 0$.

\newpage

\subsection{Non-vanishing cosmological constant}

Let us now turn our attention to the case of $\L > 0$.
In the following plots the curves correspond to values 
of $\L$ with $\ln(\L)\in\{+3, 0, -3, -6, -9, -12, -15\}$ (remember that we are 
using the normalization conventions of Secs.~\ref{ivp}, \ref{subsecpd}).
Since current data seem to indicate that $Ht$ is very close to 1 presently,
we are following the evolution of our universes only up to the point where
$Ht$ exceeds $3/2$; this is the reason why the various curves seem to end
prematurely.

\begin{figure}[H]
\begin{center}
\includegraphics[width=12cm]{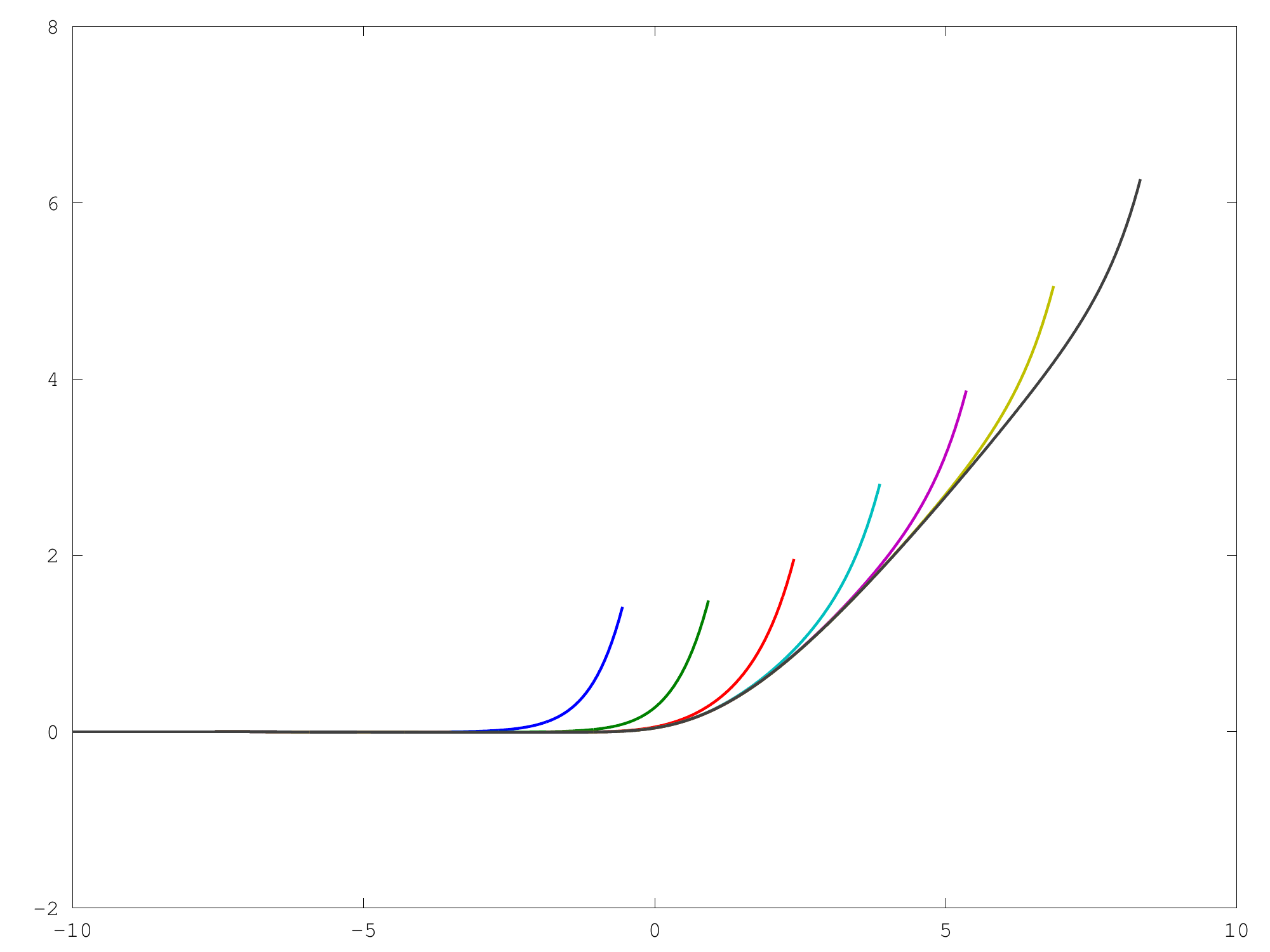}
\capt{12cm}{eu87lvot2}{ln($V_\cd/t^2$) over ln($t$) for 
$\ln(\L) = +3/0/-3/-6/-9/-12/-15$}
\end{center}
\end{figure}

\fref{eu87lvot2} shows again ln($V_\cd/t^2$) plotted over ln($t$).
There is no visible difference between the curves for $\ln(t) < -3$, and the 
whole graph looks like a single curve (which would, of course, just be the curve
corrseponding to an inhomogeneous universe with $\L = 0$) sprouting arms at 
different locations.
Let us briefly compare this with the standard FLRW case without 
inhomogeneities.
In that case the same figure would look just like a straight line
sprouting arms (we have checked that with our programs); 
similar statements hold for the next two figures.

\begin{figure}[H]
\begin{center}
\includegraphics[width=12cm]{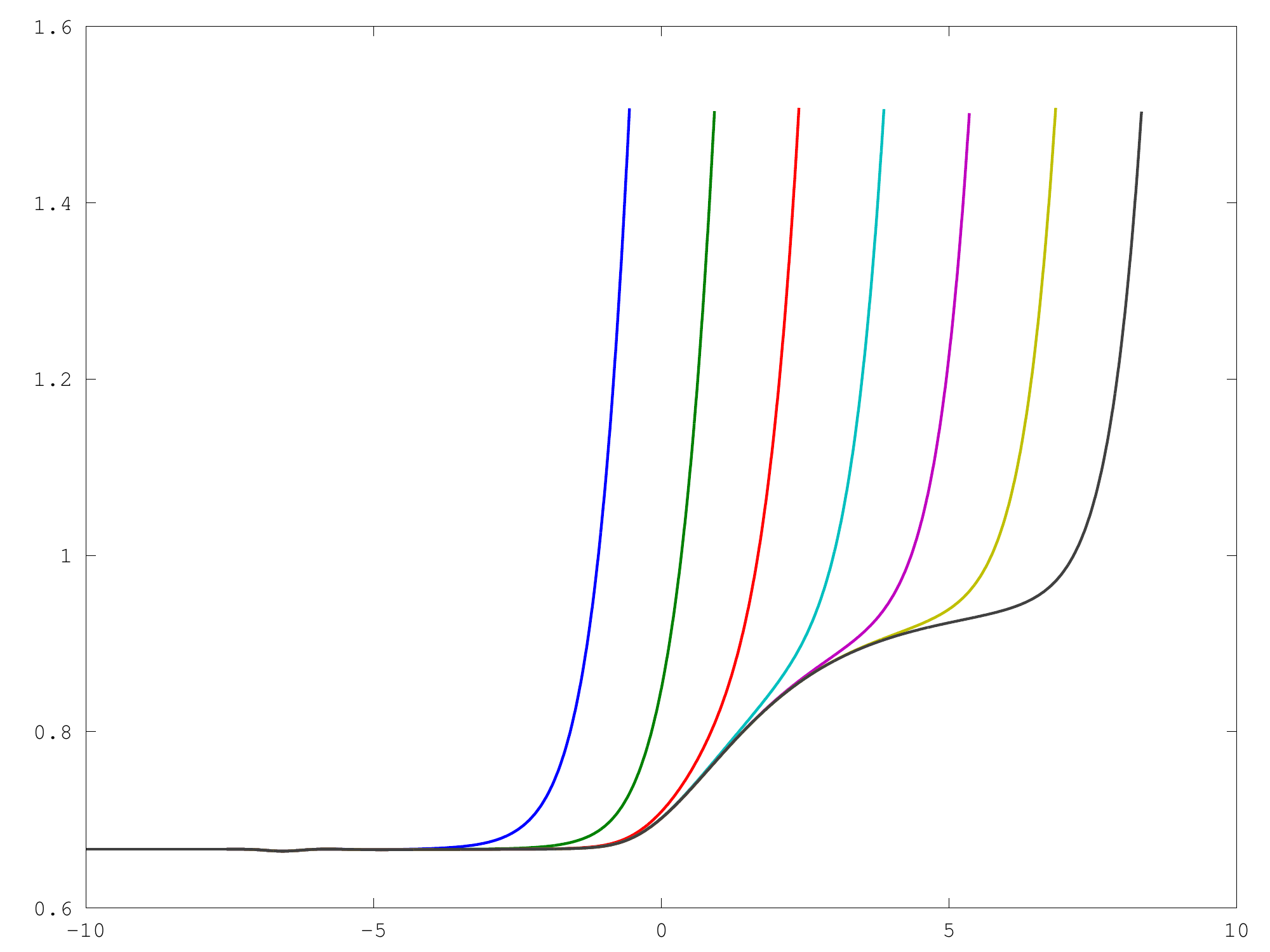}
\capt{12cm}{eu87Ht}{$Ht$ over ln($t$) for $\ln(\L) = +3/0/-3/-6/-9/-12/-15$}
\end{center}
\end{figure}

\begin{figure}[H]
\begin{center}
\includegraphics[width=12cm]{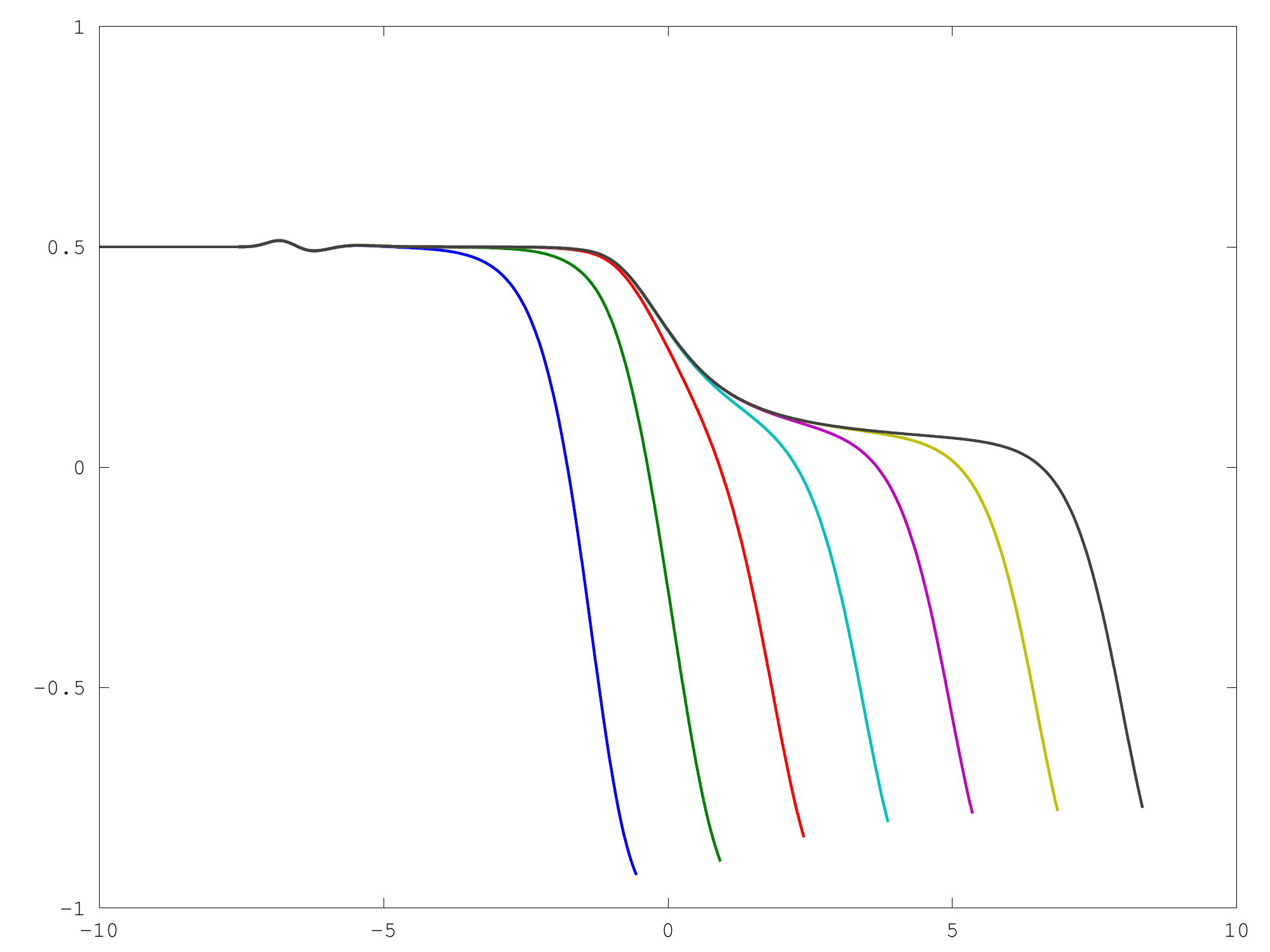}
\capt{12cm}{eu87decel}{Deceleration parameter over ln($t$) for 
$\ln(\L) = +3/0/-3/-6/-9/-12/-15$}
\end{center}
\end{figure}

Figs.~\ref{eu87Ht} and \ref{eu87decel} again show $Ht$ and the deceleration 
parameter plotted over ln($t$) for our chosen values of $\L$.
The wiggle near $\ln(t)\approx -6$ is a numerical artefact that should be 
ignored.

\begin{figure}[H]
\begin{center}
\includegraphics[width=12cm]{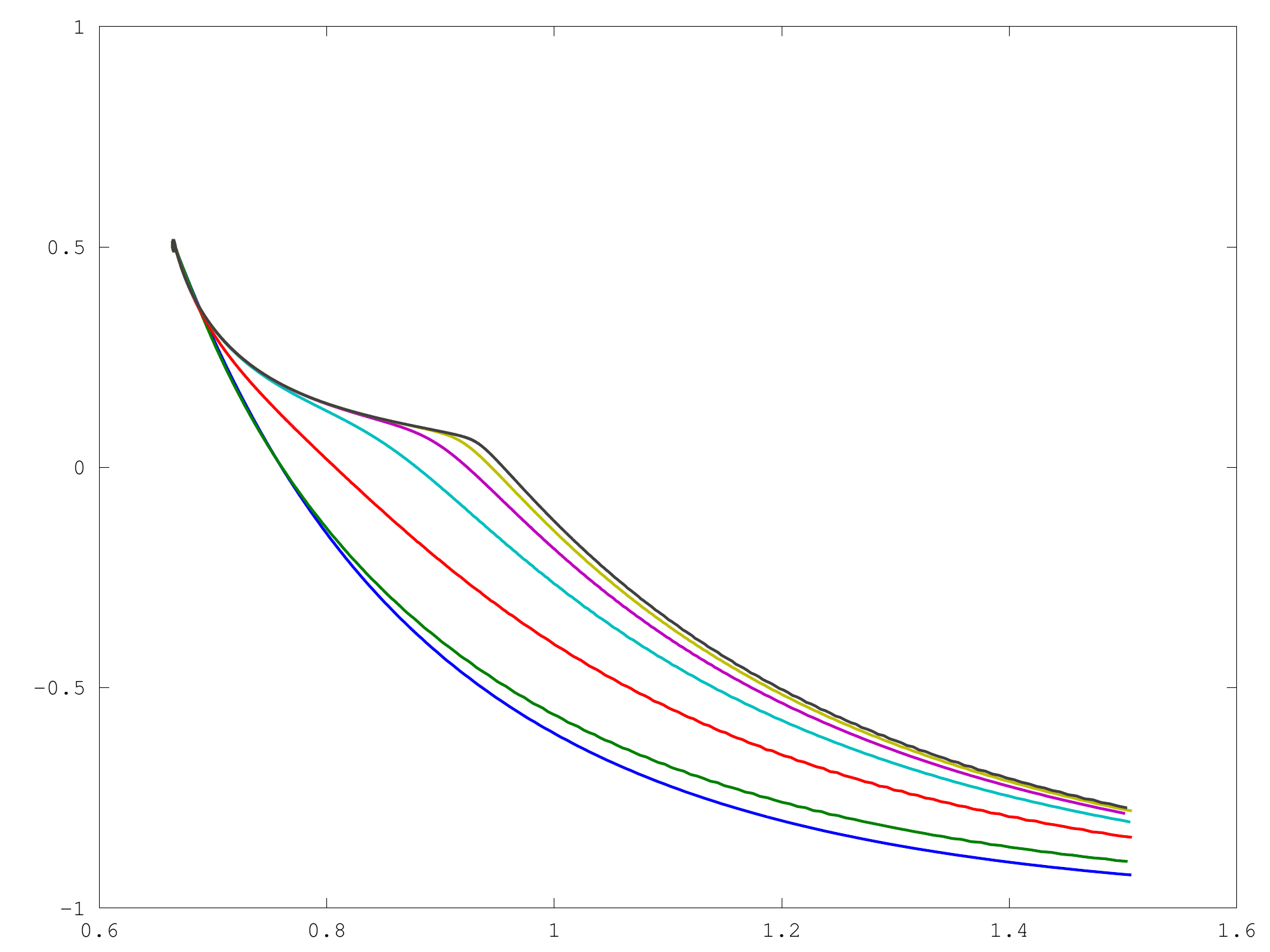}
\capt{12cm}{eu87qHt}{Deceleration parameter over $Ht$ for 
$\ln(\L) = +3/0/-3/-6/-9/-12/-15$}
\end{center}
\end{figure}

\fref{eu87qHt} is the analog of \fref{beu7qHt} for the case of $\L \ne 0$,
displaying again the deceleration parameter $q$ over $Ht$.
In contradistinction to the case of $\L = 0$, it is now
possible to 
produce the currently observed values for $Ht$ and $q$.
This requires being close to the green (second lowest) curve which corresponds
to a value of $\ln(\L) = 0$ in our dimensionless units.
By taking another look at Figs.~\ref{eu87Ht} and \ref{eu87decel} we can 
identify the time when these values are taken (i.e., today) as corresponding
to $\ln(t)\approx 0$ in our dimensionless units.
We stress the fact that 
the occurrence of the
values $\L\approx 1$, $t\approx 1$ 
was \emph{not} built into the model but emerged as a surprising result.

\newpage

\section{Discussion}\label{secdisc}
Let us start the discussion of our results with the observation
that we have an \emph{extremely predictive} model:
once values for the cosmological constant $\L$ and the parameter $S_b$
controlling the background curvature have been chosen, the evolution of
the volume of the universe is completely fixed.
In order to compare the results of the model with observations, the 
only quantity that still needs to be determined is the scale of the time
parameter. 

Inhomogeneity strongly affects the evolution of the universe, but it does 
not predict a negative deceleration parameter in a flat universe with $\L = 0$.
If we assume that cosmological observations ``see'' the quantities $H_\cd$
and $q_\cd$ that we have computed (but see below for 
a brief discussion of this assumption),
then \fref{eu87qHt} suggests that we should choose $\L\approx 1$ 
and that the present age of the universe would correspond to $t\approx 1$. 
These statements refer to our dimensionless units where $t\approx 1$ is just
the time when the inhomogeneities become relevant, and $\L\approx 1$ is the 
value for which the effects of $\L$ become strong precisely at that time.
If this conclusion holds, it appears like yet another remarkable 
coincidence.

Another possibility to calibrate our model would be to match it
with the cosmic microwave background, where the deviations from uniformity
that form the basis of our model have been measured to great precision.
A comparison is not completely straightforward, however, since the quantity 
$\D_\cR^2$ that parametrizes perturbations is usually quoted in terms of its
Fourier modes, whereas in our approach we have integrated over them.
For a comparison with $\D_\cR^2$ we would have to evaluate the integral
(\ref{Iint}) completely, not just up to the constant $I$; unfortunately 
this integral would be infinite with the standard scale--invariant spectrum, 
so we would have to think carefully about the appropriate cutoff.
Alternatively one could try to match the data directly to some of our
formulas, such as Eq.~(\ref{Deltarho}).

In order to assess the validity of our results, let us reiterate the 
simplifying assumptions that were made.

The simplification that is most relevant to our model is the
irrotational dust approximation.
We know that this approximation breaks down both in the early 
universe and in regions that have virialized after contracting.
As to the early universe, there certainly exist many observable effects
(e.g.~baryogenesis, baryon acoustic oscillations etc.) which cannot be
explained within the irrotational dust setup.
Clearly we should not apply our model to the era before matter domination.
But matter domination starts approximately at the time of decoupling,
when linear perturbation theory still provides an excellent description
of the physics of the universe.
Since our only assumptions on initial values are compatibility with 
linear perturbation theory and independence of the different Fourier modes, 
the breakdown of the irrotational dust approximation in the early universe 
should not lead to problems for our model.
Regarding the treatment of 
collapsed regions, perhaps the simplest approach to estimating the effect of 
the ambiguity in modelling collapse is to directly compare different ways of 
treating collapsing regions.
This can easily be done, with the result that this ambiguity seems to play
a very minor role. For example, at $\L = 0$, $S_b = 0$, $t=10$ we get
$V_\cd = 126.62$ if we simply assume that there is no collapse at all, 
i.e.~that any region that should collapse remains at its maximum volume;
we get $V_\cd = 124.41$ if we assume that collapse is stopped at half the
maximal extension (as we did for the results we presented in the previous 
section); 
and we get $V_\cd = 124.17$ if we allow such regions 
to collapse completely (i.e.~to size zero).
For smaller $t$ the differences are even smaller.

\begin{figure}[H]
\begin{center}
\includegraphics[width=12cm]{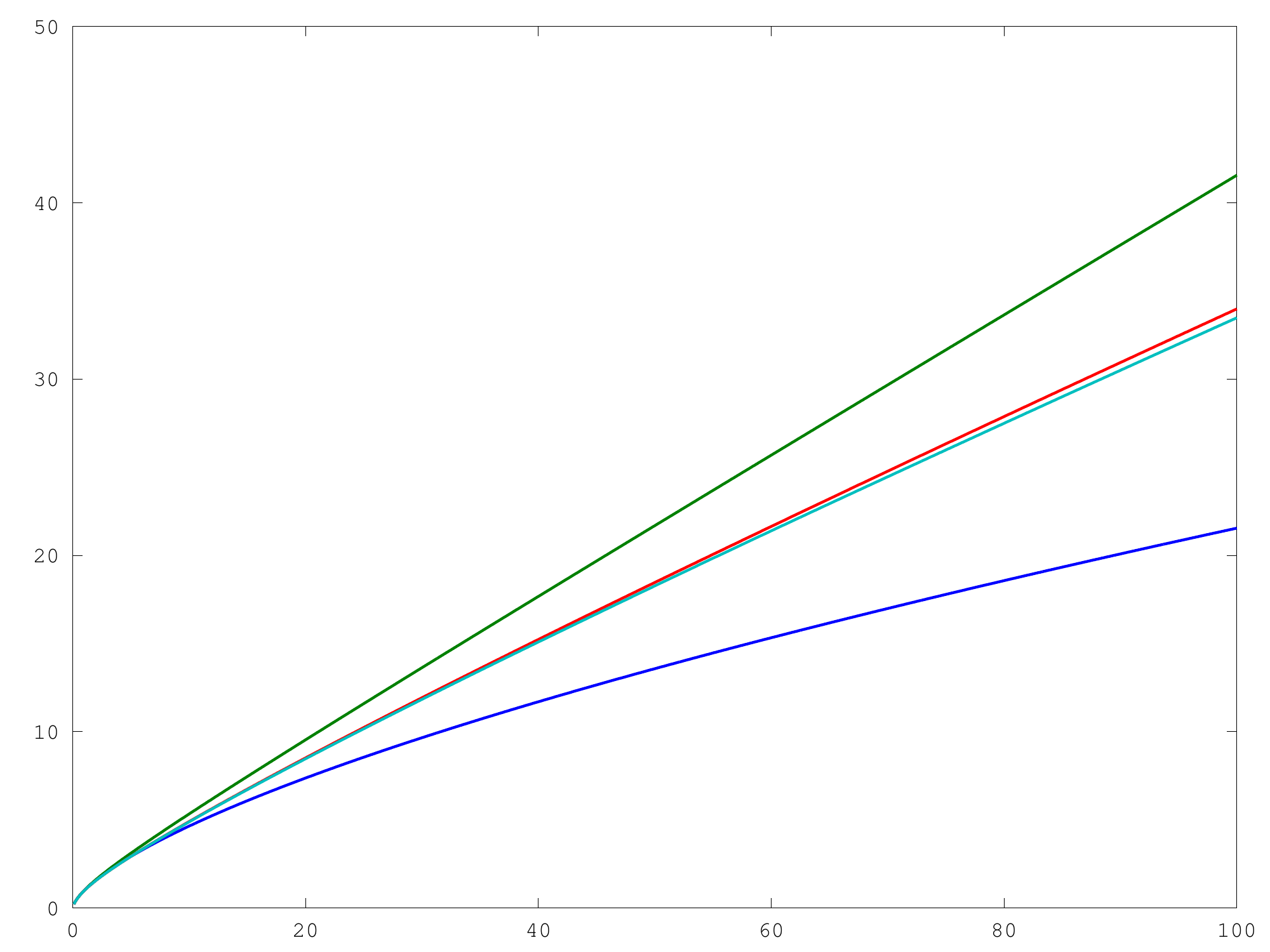}
\capt{12cm}{a100}{$a_\cd$ plotted over $t$ for an EdS universe and versions
1.-3. of our model as presented in Sec.~\ref{ivp}, with $\L = 0$ and $S_b = 0$}
\end{center}
\end{figure}

Another, possibly important source of imprecision is the treatment of 
the traceless part of the Ricci tensor.
We discussed various possibilities in Sec.~\ref{ivp} around 
Eq.~(\ref{rbarijev}).
While we cannot simulate the exact behaviour including what we called the 
$Y$-terms, we have performed the computations for the cases 1.-3. and compared
them with the homogeneous EdS case.
\fref{a100} displays the results, with each of the lines corresponding to
a plot of $a_\cd$ over $t$ for $\L = 0$ and $S_b = 0$, for one of the 
following scenarios. 
The lowest (blue) line corresponds to an EdS universe.
The highest (green) line corresponds to our first scenario, i.e.~it takes 
into account the inhomogeneities, but ignores
the impact of shear and the traceless part of the Ricci tensor as quantified 
by the last term in Eq.~(\ref{lsfev}).
The second highest (red) line indicates the result of scenario 2., where 
the inhomogeneous part $\bar r$ of the rescaled Ricci-tensor is modelled as 
constant. 
The line immediately below (shown in cyan) indicates scenario 3., 
i.e.~the result of computing the evolution of $\bar r$ via 
Eq.~(\ref{rbarijev}).
Note that all the lines would be much closer if we had chosen logarithmic 
scales as in the previous section.
While it is obviously important to include the inhomogeneities and 
a nonzero value of $\bar r$, 
it does not seem to make much difference whether we take that quantity to 
be constant or whether we take into account those aspects of its
evolution that are captured by Eq.~(\ref{rbarijev}).
Therefore one would hope that the omission of the $Y$-terms should play a 
similarly small role; nevertheless, this is perhaps the weakest point of
the present model.

A minor source of uncertainty lies in our application of the
results of Ref.~\cite{LS}. 
Our assumption that only the function $C(x)$ is relevant relies on
ignoring tensor modes and decaying modes.
It is very hard to see how this might affect our results in a serious manner.

Given the computational nature of the present results, one should also take
into account the possibility of mistakes in programming and the effects of
numerical imprecisions.
It seems unlikely that such effects should modify the general conclusions
of this work, for the following reason.
During the programming phase, several mistakes were made and later eliminated,
and in each case there was a change in numerical details of the results, but not
in the overall structures.
Therefore, even if there are still some factors of 2 or $\pi$ wrong in the 
programs, we would not expect a significant modification of the general 
conclusions.

So, have we shown that a cosmological constant or dark energy is indeed needed
to account for the results of observations?
There may still be one important loophole: in our attempts to match the results
of our computations with data, we have assumed that the various observational
devices actually ``see'' the quantities $H_\cd$ and $q_\cd$ that we defined
by volume averages in our inhomogeneous model universes.
But there is no such thing as a device that can measure cosmological volumes 
directly.
Instead, all the data 
come from photons that have travelled to us through an inhomogeneous universe. 
If light propagation in such a case is essentially equivalent to light 
propagation in a homogeneous universe with the scale factor $a_\cd$,
then there seems to be no escape from having to assume $\L > 0$;
otherwise it would be possible 
that acceleration is simulated without taking place.
In either case the methods presented here should provide useful tools for
interpreting the results of precision cosmology.

\del
\section{Further possibilities}
Collapse everyhwere: $\bar \s > a\,t\, r_\mathrm{in}/2$ implies
\beq 
a^{-5}\bar \s^2 > {1\0 4}r_\mathrm{in}^2 a^{-3}t^2 > \hbox{const} \times t^{-1} 
\eeq
$\bar\s - \bar r$ system: attractor to $(1, -\2, -\2)$\\
percentages of matter in expanding / collapsing / collapsed regions\\
percentages of volumes in expanding / collapsing / collapsed regions\\
\enddel

{\it Acknowledgements:} 
It is a pleasure to thank Anton Rebhan and Dominik Schwarz for
helpful discussions.

\section*{Appendix: Details of the computation}

The actual computations were performed numerically with the help of GNU octave
\cite{GNU}.
The (integro-)differential equations (\ref{a32ev}) to (\ref{rbarijev}) were
discretized explicitly, using the Euler method.
Two different approaches were implemented: on the one hand, equal steps 
$\epsilon = \bar t_{n+1} - \bar t_n$ were used to model the time parameter
$\bar t$ itself; on the other hand $\ln \bar t$ was subjected to an equidistant
discretization, resulting in a constant value for $\bar t_{n+1} / \bar t_n$.
While the first (equal $\bar t$-steps) approach is more straightforward 
it is not so successful in combination with
the $\bar S$-- and $\bar \d$--integrations in parameter space, where the 
finiteness of the range of $\bar t$ values quickly leads to problems.
Here the second (equal ln--steps) approach, where even a moderate number of 
values $\bar t_n$
can span a reasonable range of orders of magnitude, is much more useful.
Therefore only the second approach was applied to the cases with 
$\L \ne 0$ or $S_b \ne 0$, and all of the figures shown in 
Sec.~\ref{secresults} were produced in this way.
However, the equal $\bar t$-steps approach was used for comparing different
ways of modelling collapse and for studying the impact of various assumptions
on the traceless part of the Ricci tensor (\fref{a100}).
For $\L = 0$,  $S_b = 0$ the results from 
both routines were compared and found to agree 
up to differences that
would be expected as errors coming from the discretization.
These two approaches were implemented in the following ways.
\subsection*{Equal $\bar t$-steps}
With this approach only the case of $\L = 0$, $S_b = 0$ was studied.
We found it useful to choose 
\beq 
q = \sqrt{{\bar S^2\0 5}+\bar \d^2},\quad
\bar S = \sqrt{5}\, q \cos\dh, \quad \bar \d = q \sin\dh,
\eeql{qforint}
so that Eq.~(\ref{result}) becomes
\bea
V_\cd(\bar t) & \sim & 
\int \bar a^3(\bar t; \sqrt{5}\, q \cos\dh, q \sin\dh, \ph) 
\myexp{-{q^2\0 2}}q^5\, \sin^4\dh \, \sin(3 \ph) dq\,d\dh d\ph\\ 
& \sim & 
\int \bar a^3(q^{3\0 2}\bar t;\sqrt{5}\,\cos\dh,\sin\dh, \ph) 
\myexp{-{q^2\0 2}}q^2\, \sin^4\dh \,     \sin(3 \ph) dq\,d\dh d\ph\\
& \sim & 
{1\0 \bar t^2}\int \bar a^3
(\bar t'; \sqrt{5}\,\cos\dh, \sin\dh, \ph) 
\myexp{-{1\0 2}({{\bar t'}\0 {\bar t}})^{4\0 3}}
      {\bar t'}\, \sin^4\dh \, \sin(3 \ph)   d{\bar t'}\,d\dh d\ph,
\eea
where we first used Eq.~(\ref{qU}) and then performed a 
change of variables from $q$ to ${\bar t'} =  q^{3\0 2}\bar t$.
In evaluating the last expression numerically, we started with the 
$\ph$--integration (whenever it was applied; see below), proceeded with 
the $\dh$--integration and finally performed the $\bar t'$--integration.
By keeping only the results of the integrations, the required memory could
be kept very small.
While the first two integrations worked well, the last one showed good
convergence properties only for a limited range of $ \bar t$--values.
This approach was used to compare the different scenarios for modelling
$\bar r$ as presented in the paragraph around Eq.~(\ref{rbarijev}), 
with the result that, while there are great differences between a homogeneous
universe and the universes corresponding to the scenarios 1.~and 2., the
difference between the scenarios 2.~and 3.~(the latter being the only one 
requiring the $\ph$--integration) were quite small.
\del
On the other hand, different versions of modelling collapse were implemented
only in the equal $\bar t$-steps approach
equal $\bar t$-steps approach was implemented only for the case of 
\enddel
%
\subsection*{Equal steps for ln($\bar t$)}
The complete range $(0,\infty)$ for $\bar t$ was divided into three parts.
While the solution was performed numerically in the middle part,
an approximation by linear perturbation theory was used for $\bar t <\!\!< 1$
and an approximation by a closed formula for 
$\bar t \in [\bar t_{\mathrm{final}}, \infty)$.
The choice of formula depended on the context:
for collapsing solutions, $\bar t_{\mathrm{final}}$ was the time step at which the
solution for $\bar a$ would have dropped below half of the maximum value
$\bar a_\mathrm{max}$, and $\bar a (\bar t)$ was taken to be 
$\bar a_\mathrm{max}/2$ for $\bar t \ge \bar t_{\mathrm{final}}$;
in the non-collapsing case $\bar a^3$ was modelled as the de Sitter solution
$\mathrm{const} \times \exp(\sqrt{3\L}\,t)$ for $\L > 0$, and
as a cubic function of $\bar t$ otherwise, with $\bar t_{\mathrm{final}}$ 
chosen in such a way that these approximations were sufficiently good.
Since the results of the equal $\bar t$-steps approach indicated that
the contribution of the term (\ref{rbarijev}) was not significant, it was 
omitted here, resulting in the simplification that no $\ph$--integration
was required.
While we did not consider the case with both $\L$ and $S_b$ non-vanishing,
we did study the following two scenarios.
\begin{itemize}
\item Flat background with a cosmological constant:\\
Parametrizing $\bar S$ and $\bar \d$ as in (\ref{qforint}) we find
\bea
V_\cd(\bar t;\bar \L) & \sim & 
\int \bar a^3(\bar t; \sqrt{5}\, q \cos\dh, q \sin\dh, \bar \L) 
\myexp{-{q^2\0 2}}q^5\, \sin^4\dh \, dq\,d\dh\\ 
& \sim & 
\int \bar a^3(q^{3\0 2}\bar t;\sqrt{5}\,\cos\dh,\sin\dh, q^{-3}\bar \L) 
\myexp{-{q^2\0 2}}q^2\, \sin^4\dh \, dq\,d\dh . \label{Lambdaint}
\eea
Again it is useful to perform the $\dh$--integration first.
In this way we generated a list of $\dh$--integrated solutions for different
values of $\bar\L$, with steps in ln($\bar\L$) twice the size of the 
ln($\bar t$)--steps.
Then computing $V_\cd(\bar t;\bar \L)$ numerically according to 
Eq.~(\ref{Lambdaint}) corresponded to just one ``diagonal'' 
(rising $\bar t$, falling $\bar \L$) summation over this list.
\item Curved background without a cosmological constant:\\
Now taking $q=\bar \d$ in Eq.~(\ref{qU}) we arrive at
\bea
V_\cd(\bar t;\bar S_b) & \sim & 
\int \bar a^3({\bar t};\bar S,\bar \d) 
\myexp{-{1\0 10}(\bar S-\bar S_b)^2-{1\0 2}\bar \d^2}\bar \d^4\, 
    d\bar S\,d\bar \d\\
 & \sim & 
\int \bar a^3(\bar \d^{3\0 2}{\bar t};{\bar S\0\bar \d}, 1) 
\myexp{-{1\0 10}(\bar S-\bar S_b)^2-{1\0 2}\bar \d^2}\bar \d\, 
    d\bar S\,d\bar \d . \label{dbarint}
\eea
This was evaluated by first generating a list of solutions 
$\bar a(\bar t, \bar S, 1, 0)$, then performing the $\bar\d$--integration
of Eq.~(\ref{dbarint}) which resulted in another list indexed by different
$\bar S$--values, and finally integrating over $\bar S$.
\del
\\ & \sim & 
{1\0 \bar t^2}\int \bar a^3({\bar t'};\bar S',1,\ph, 0) 
\myexp{-{1\0 10}\(({{\bar t'}\0 {\bar t}})^{2\0 3}\bar S'-\bar S_b\)^2
   -{1\0 2}({{\bar t'}\0 {\bar t}})^{4\0 3}}\bar t'\, 
    \sin(3 \ph) d\bar S'\,d\bar t'\, d\ph.\eea
\enddel
\end{itemize}
The numerical parameters (step widths, values at which the 
description changes from perturbation theory to explicit computation, 
etc.)~were chosen pragmatically in such a way that a further
refinement would not lead to clearly discernible effects in the plots;
an exception is the wiggle in \fref{eu87Ht} which could have been reduced
only by a very time consuming increase in the number of steps.
For producing the plots of Sec.~\ref{secresults} we took the stepwidth in 
ln($t$) to be $2^{-7}$, divided the interval $[0,\pi]$ for $\dh$ into $2^8$ 
parts, created a list of solutions for values $\ln(\L)\in[-27,9]$ (with a
stepwidth of $2^{-6}$, i.e.~twice the value of that for $\ln(t)$), 
and similarly chose further parameters.

The differentiations required to compute $H$ and $q$ were performed numerically
with formulas such as 
\beq H(t_n)\,t_n = {d\ln(a) \0 d\ln(t)} (t_n) \approx 
{\ln(a_{n+1}) - \ln(a_{n-1}) \0  \ln(t_{n+1}) - \ln(t_{n-1})}.
\eeq
This worked quite well in general.
The exceptions were the aforementioned wiggle, as well
as a spike that would have occurred in Figs.~\ref{beu7decel} and 
\ref{beu7deceltom4} as a consquence of the transition between different 
regimes; its artificiality was clear from
the fact that it became larger when the stepwidth was refined.
For producing the figures we simply removed the two offending
data points by hand.



\bye